\documentclass{article} 
\usepackage{iclr2024_conference,times}

\usepackage{amsmath,amsfonts,bm}

\def\eqref#1{equation~\ref{#1}}

\def\1{\bm{1}}

\DeclareMathAlphabet{\mathsfit}{\encodingdefault}{\sfdefault}{m}{sl}
\SetMathAlphabet{\mathsfit}{bold}{\encodingdefault}{\sfdefault}{bx}{n}

\usepackage{caption}
\usepackage{subcaption}
\usepackage{appendix}
\usepackage{bbm}
\usepackage{hyperref}
\usepackage{url}
\usepackage{gensymb}
\usepackage{graphicx}
\usepackage{caption}
\usepackage{multirow}
\usepackage{makecell}
\usepackage[frozencache,cachedir=.]{minted}
\usepackage[minted,most]{tcolorbox}
\usepackage{xcolor} 
\definecolor{LightGray}{rgb}{1,0.98,0.85}
\usepackage{listings}
\usepackage{cleveref}
\usepackage[T1]{fontenc} 
\usepackage{wrapfig}
\usepackage{booktabs} 

\usepackage{titlesec}
\titlespacing*{\subsection}
{0pt}{0.9ex}{0.7ex}
\newcommand{\python}[3]{
\noindent\rule{\linewidth}{0.3pt}
\vspace{-2em}
\inputminted[linenos, breaklines, fontsize=\footnotesize, numbersep=5pt, fontsize=\scriptsize, xleftmargin=12pt]{python}{#1}
\vspace{-1.5em}
\noindent\rule{\linewidth}{0.3pt}

\noindent\begin{minipage}{\linewidth}
\captionof{listing}{#3}
\label{#2}
\end{minipage}
\vspace{-8pt}
}

\newcommand{\prompt}[3]{
\noindent\rule{\linewidth}{0.3pt}
\inputminted[linenos, breaklines, fontsize=\footnotesize, numbersep=5pt, fontsize=\scriptsize, xleftmargin=15pt]{text}{#1}
\noindent\rule{\linewidth}{0.3pt}

\noindent\begin{minipage}{\linewidth}
\captionof{listing}{#3}
\label{#2}
\end{minipage}
\vspace{-8pt}
}

\title{LangProp: A code optimization framework using Large Language Models applied to driving}

\author{
\bf{Shu Ishida$^{1,2}$\thanks{Work done during an internship at Wayve}, Gianluca Corrado$^1$, George Fedoseev$^1$, Hudson Yeo$^1$,}\\
\bf{Lloyd Russell$^1$, Jamie Shotton$^1$, Jo\~{a}o F. Henriques$^2$ \& Anthony Hu$^1$}\\
$^1$Wayve Technologies \quad $^2$Visual Geometry Group, University of Oxford\\
\texttt{research@wayve.ai, ishida@robots.ox.ac.uk}
}

\iclrfinalcopy 
\begin{document}

\maketitle

\begin{abstract}
We propose LangProp, a framework for iteratively optimizing code generated by large language models (LLMs), in both supervised and reinforcement learning settings. While LLMs can generate sensible coding solutions zero-shot, they are often sub-optimal. Especially for code generation tasks, it is likely that the initial code will fail on certain edge cases. LangProp automatically evaluates the code performance on a dataset of input-output pairs, catches any exceptions, and feeds the results back to the LLM in the training loop, so that the LLM can iteratively improve the code it generates. By adopting a metric- and data-driven training paradigm for this code optimization procedure, one could easily adapt findings from traditional machine learning techniques such as imitation learning, DAgger, and reinforcement learning. 
We show LangProp's applicability to general domains such as Sudoku and CartPole, as well as demonstrate the first proof of concept of automated code optimization for autonomous driving in CARLA. We show that LangProp can generate interpretable and transparent policies that can be verified and improved in a metric- and data-driven way. 
Our code is available at \url{https://github.com/shuishida/LangProp}.
\end{abstract}

\section{Introduction}
Building systems that can self-improve with data is at the core of the machine learning paradigm. 
By leveraging vast amounts of data and having an automated feedback loop to update models according to an objective function, machine learning methods can directly optimize the metrics of interest, thus outperforming systems that are handcrafted by experts. In the early history of artificial intelligence (AI), Symbolic AI, e.g. rule-based expert systems~\citep{hayes1985rule,jackson1986introduction}, was a dominant and perhaps a more intuitive and explainable approach to solving tasks in an automated way, and is still widely used in fields such as medicine~\citep{abu2017medical} and autonomous driving~\citep{badue2021self}. However, there have been numerous successes in recent decades in machine learning, e.g. deep neural networks, that demonstrate the advantage of data-driven learning.

Advances in Large Language Models (LLMs)~\citep{NEURIPS2020_gpt3,openai2023gpt4,touvron2023llama} were enabled by neural networks. Trained on both natural language and code, they can translate human intent and logic into executable code and back, expanding the boundaries of applying logic and reasoning. Unlike other machine learning techniques, LLMs have an affinity with Symbolic AI since they operate in discrete symbolic input-output spaces. The generated outputs are interpretable, even though the internal representation of these tokens is in a continuous embedding space. 
This observation led us to question if it is possible to have the best of both worlds -- having an interpretable and transparent system, characteristic of Symbolic AI, which can self-improve in a data-driven manner, following the machine learning paradigm. We believe that LLMs provide the missing piece of the puzzle; the optimization mechanism. 

Our insight is that we can draw a direct analogy from training neural networks, and \emph{train} symbolic systems by leveraging the power of LLMs to interpret and generate scripts.
Using this analogy, an LLM can be considered as an \emph{optimizer} equivalent to stochastic gradient descent or Adam. The actual \emph{model} in our paradigm is an object that handles the initialization and updates of \emph{parameters} as well as the forward pass logic, where the \emph{parameters} are a collection of symbolic scripts that the LLM generates. At every iteration, we perform a forward pass through the model, compare it against the ground truth in the dataset, and pass the scores and feedback into the LLM which interprets the results and updates the scripts in a way that fixes the issues raised.

While many methods use LLMs for code generation, and systems such as Auto-GPT~\citep{autogpt} iteratively query LLMs to execute tasks in an agent-like manner, as far as we know, we are the first to completely translate and apply the training paradigm used in machine learning for iterative code generation. We draw inspiration from \textsc{Voyager}~\citep{wang2023voyager}, which introduced the idea that a collection of LLM-generated code (skill library) can be considered as sharable and fine-tunable \emph{checkpoints}. However, \textsc{Voyager}'s method is specific to Minecraft, and additional work is needed to apply its approach to other domains. We propose LangProp, a code optimization framework that is easily adaptable to many application domains.


Autonomous driving is a key area in which model interpretability and transparency are critical. We consider LangProp to be a valuable proof of concept for building interpretable and language-instructable systems in a more automated and learnable way.
We tested our hypotheses that (a) LangProp can generate interpretable code that learns to control a vehicle, (b) LangProp can improve driving performance with more training data in comparison to zero-shot code generation, and (c) we can easily transfer training paradigms from machine learning to LangProp such as imitation learning (IL), reinforcement learning (RL)~\citep{sutton2018reinforcement} and DAgger~\citep{ross2011reduction_dagger}.

\section{Related work}
\subsection{LLMs for code generation}
Transformers ~\citep{vaswani2017attention} have shown outstanding performance in code generation tasks~\citep{chen2021evaluating_codex,li2022competition_alphacode,xu_polycoder2022,nijkamp2022codegen,fried2022incoder}. In particular, general purpose LLMs~\citep{NEURIPS2022_b1efde53_instructgpt_human_feedback, openai2023gpt4} have shown remarkable capabilities of translating between natural language and code. However, there is no guarantee that the generated code is error-free. Benchmarks have been suggested to evaluate LLMs on the code generation quality~\citep{chen2021evaluating_codex,liu2023_isyourcodecorrect}. 
Code generation with execution is highly relevant to our work. \citet{cobbe2021training} and \citet{li2022competition_alphacode} used majority voting on the execution results to select code from a pool of candidates. but this is prone to favoring common wrong solutions over correct solutions. \citet{pmlr-v202-ni23b-lever} suggested a ranking mechanism using a learned verifier to assess code correctness. 
\textsc{CLAIRify}~\citep{skreta2023errors} implemented automatic iterative prompting that catches errors and provides feedback to the LLM until all issues are resolved. 

Tangentially related fields are Automated Program Repair~\citep{xia2022less_coderepair,xia2022practical_coderepair}, unit test generation~\citep{roziere2021leveraging_unittests}, and planning for code generation \citep{le2022coderl,zhang2022planning}. 
While orthogonal to our approach of iteratively generating code using a pre-trained general-purpose LLM as an optimizer, findings from these fields may be compatible with LangProp.

\subsection{LLMs for automating compositional tasks}
LLM-powered agents have demonstrated sophisticated planning capabilities. Sequential prompting with the history of observation, action, and the reason for the action was proposed by ReAct~\citep{yao2022react} as an improvement to Chain-of-Thought prompting~\citep{wei2022chainofthought}.
Auto-GPT~\citep{autogpt} automated tasks by iteratively generating a sequence of subtasks in finer detail until they are executable. 
SayCan~\citep{ahn2022saycan} used LLMs to generate candidate subgoals and assessed their affordances with a value function given visual observations to ground the agent's behavior.
VIMA~\citep{jiang2023vima} and PaLM-E~\citep{driess2023palme} demonstrated profound reasoning and execution capabilities on multi-modal tasks such as Visual Q\&A and robotics by fine-tuning LLMs to allow multi-modal prompting.
Inner Monologue~\citep{huang2023inner_monologue} used environment and user feedback to replan for embodied tasks. Unlike our method, the above methods require an LLM in the loop during inference, whereas our method only requires access to an LLM during the code optimization stage. \citet{liang2023codeaspolicies} and \citet{singh2023progprompt} used LLMs to directly generate code for robotics, while ViperGPT~\citep{surismenon2023vipergpt} and VisProg~\citep{gupta2023visprog} composed pre-trained vision-and-language models to solve challenging vision tasks which require reasoning and domain knowledge.
However, none of the above methods implement code optimization via iterative prompting.

Our method is inspired by \textsc{Voyager}~\citep{wang2023voyager}, which integrates environment feedback, execution errors, and self-verification into an iterative prompting mechanism for embodied control in Minecraft. \textsc{Voyager} maintains a \emph{skill library}, a collection of verified reusable code, which can be considered as \emph{checkpoints}. However, there is no mechanism to optimize or remove a sub-optimal skill in the skill library. We address this limitation and present a more general code optimization framework that can be applied to a variety of domains, including autonomous driving.

\begin{figure}[t]
\includegraphics[width=\textwidth]
{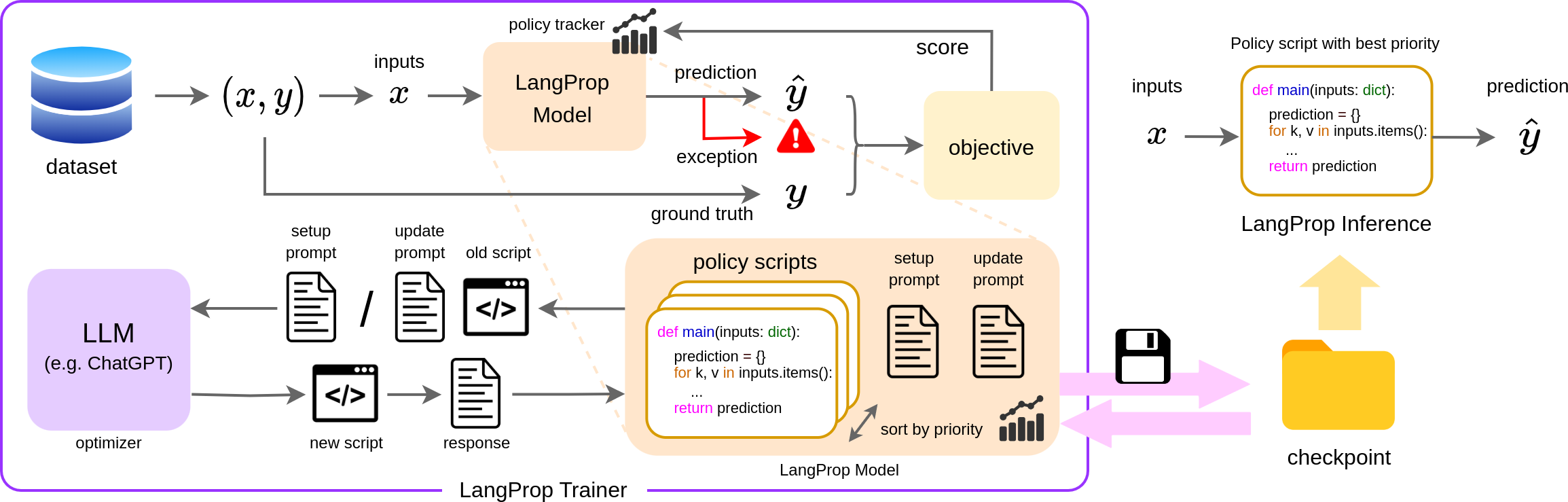}
\caption{An overview of the LangProp framework, which consists of a LangProp model, an LLM optimizer, and a LangProp trainer. During training, the LLM generates and updates the policy scripts which are evaluated against a training objective. The performances of the policies are monitored and aggregated over time by a policy tracker as \emph{priorities}, which is then used to rerank the policies. 
Policies with higher priorities are selected for updates, and the best policy is used for inference.
}
\label{fig:langprop_overview}
\vspace{-8pt}
\end{figure}

\section{The LangProp Framework}
\label{sec:langprop}
The LangProp framework, shown in \Cref{fig:langprop_overview}, addresses a general task of optimizing code on a given metric of success in a data-driven way, similar to how a neural network is optimized on an objective function. LangProp performs iterative prompting to improve code performance, using the inputs, outputs, exceptions, metric scores, and any environmental feedback to inform the LLM upon updates. The updates in LangProp are performed using a form of an evolutionary algorithm~\citep{back1993overview_evolution}. The following sections describe the key concepts in LangProp in more detail.

\subsection{Model definition}
The LangProp model consists of a setup prompt, an update prompt, and a collection of executable code generated by the LLM, which we refer to as \emph{policies}. While neural models are parameterized by floating-point weights, the \emph{parameters} of a LangProp model is the set of policies. Each policy is associated with an executable \emph{script} as well as a statistics tracker, which updates the \emph{priority}, an aggregate measure of the policy's performance with respect to the training objective. The priority is used to rerank policies so that the best-performing policies are used for updates and inference. 

\subsubsection{Policy setup}
\label{sec:policy_setup}
The initialization of the policies is done similarly to zero-shot code generation.
The definition and specification of the requested function are given as a docstring of the function, including the names and types of the inputs and outputs, what the function is supposed to achieve, and a template for the function. We also adopt Chain-of-Thought prompting~\citep{wei2022chainofthought}. Examples of setup prompts can be found in \Cref{sec:setup_prompt_example}. Responses from the LLM are parsed to extract the solution code snippets. Multiple responses are collected to ensure the diversity of the initial policies.

\subsubsection{Training objective}
The difference of LangProp over typical usage of LLMs for code generation is that it performs code optimization in a metric- and data-driven manner. In many tasks, it is easier to provide a dataset of inputs and ground truth corresponding outputs rather than to accurately specify the requirements for a valid solution or write comprehensive unit tests. Similar to how neural networks are trained, the user defines an objective function that measures how accurate the policy prediction is against the ground truth, e.g. L1 or L2 loss. A penalty is given if the policy raises an exception.

\subsubsection{Forward-pass and feedback}
Similar to training neural networks, LangProp assumes a dataset of inputs and associated ground truth labels for supervised learning (or rewards/returns for reinforcement learning, discussed in \Cref{sec:agent_training}). 
For every batch update, the inputs are fed into all the policies currently in the LangProp model to make predictions, equivalent to a \emph{forward-pass}. For each policy, the prediction is evaluated by the objective function which returns a \emph{score}. If an exception is raised during execution of a policy script, it is caught by the model and an exception penalty is returned as a score instead. 

The execution results, which include the score, exception trace, and any printed messages from the execution, are fed back into the model and are recorded by the policy tracker. This is analogous to how neural network parameters are assigned gradients during back-propagation (see \Cref{sec:backpropagation}). This information stored by the tracker is used in the policy update step in \Cref{sec:policy_update}.

\subsubsection{Priority}
The priority is, simply put, an average of scores with respect to the training objective. In case a small batch size is required for faster computation, a running average of the scores is used as the priority rather than ranking the policies' performance based on scores from the current batch alone, which may result in highly stochastic results. This is sufficient for supervised learning with a fixed-size dataset. As discussed later in \Cref{sec:agent_training}, however, a more complex training method such as reinforcement learning or DAgger~\citep{ross2011reduction_dagger} has a non-stationary training distribution. Therefore, we use exponential averaging with a discount factor of $\gamma \in (0, 1]$ following \Cref{eq:exp_avg}.
{
\thinmuskip=1mu
\medmuskip=2mu
\thickmuskip=3mu
\begin{equation}
\label{eq:exp_avg}
P_{i,k} = \left(\sum_{j=1}^{N_k^B} s_{i,j,k} + W_{i,k-1} P_{i,k-1}\right) / \left(N_k^B + {W}_{i,k-1}\right), \quad {W}_{i,k} = \gamma (N_k^B + {W}_{i,k-1})
\end{equation}
}

Here, $N_k^B$, $P_{i,k}$ and $W_{i,k}$ are the batch size, priority, and priority weighting of the $k$-th batch for the $i$-th policy, respectively, and $s_{i,j,k}$ is the objective score of the $i$-th policy for the $j$-th element in the $k$-th batch. Initial conditions are $P_{i,0} = 0$ and $W_{i,0} = 0$. By weighting recent scores higher, we ensure policies with higher priorities have high performance on the most up-to-date dataset. 

\subsubsection{Policy reranking and update}
\label{sec:policy_update}
This step updates the model based on the most recent forward-backward pass and updated priorities. This corresponds to the optimization step in neural network training, where parameters are updated based on gradients computed on the most recent batch. First, the policies are reranked by the priorities and the top $N^K$ number of policies are kept, out of which the top $N^U$ policies are selected for updates. For each of these policies, the policy tracker is queried for the worst-case input-output pairs in the training batch, namely that with the lowest objective score. The tracker returns the corresponding input, output and score, along with any exception or print messages during the execution. This information, together with the old policy script, is embedded into the update prompt by a prompt template engine (\Cref{sec:template_engine}). The update prompt is passed to the LLM, which returns $N^R$ responses containing new policy scripts for each of the $N^U$ policies chosen for updates.

After the update, there are $N^U \times N^R$ new policies and up to $N^K$ old policies. To initialize the new policies with sensible priorities, objective scores for the new policies are evaluated by performing the forward-backward pass, using the same training samples as the current update. Finally, all the policies are sorted by their priorities, ready for inference or training on a new batch.

\subsection{Prompt template engine}
\label{sec:template_engine}
During the policy update stage, we require a dynamic prompting mechanism to embed information about the input, predicted output, ground truth, exception, print messages, and the policy script to be revised. The logic to generate these prompts is sometimes complex, for example, predictions are only made when there are no exceptions. To enable flexible prompt generation while avoiding any hardcoding of the prompts in the codebase, we developed a simple yet powerful prompt template that can parse variables, execute Python code embedded within the prompt, and import sub-prompts from other files, and will be included in our open-sourced solution. The update prompt examples shown in \Cref{sec:update_prompt_example} make extensive use of the policy template engine's capabilities.

\subsection{Training paradigm}

LangProp mirrors the code abstraction of PyTorch~\citep{paszke2019pytorch} and PyTorch Lightning~\citep{falcon2019pytorchlightning} for the module and trainer interfaces. This allows LangProp to be task-agnostic, making it easily applicable to a range of domains and use cases. Moreover, it helps highlight the similarities between neural network optimization and code optimization using LangProp and facilitates a smooth integration of other neural network training paradigms.

Importantly, LangProp's internal implementation does not depend on PyTorch or PyTorch Lightning. LangProp supports PyTorch datasets and data loaders, as well as any iterable dataset object for training and validation. Listing \ref{code:training} shows an example of a standard LangProp training script.

\python{code/langprop_train.py}{code:training}{Training a LangProp model with a trainer. The model can be instantiated from a path to the setup and update prompts that specify the task to be learned.}

After every training step on a mini-batch, the trainer saves a \emph{checkpoint}, which consists of the setup prompt, update prompt template, the policy scripts (maximum of $N^K + N^U \times N^R$), and the statistics monitored by the policy tracker (priorities $P$ and priority weights $W$). Since these can be stored as text or JSON files, the size of a checkpoint is in the order of a few hundred kilobytes. Checkpoints can be used to resume training, fine-tune the model, or for inference.

\python{code/langprop_inference.py}{code:inference}{Inference with a LangProp model checkpoint.}

Listing \ref{code:inference} shows how a LangProp checkpoint can be loaded and used for inference. The policy with the highest priority is used. Since policies are \emph{parameterized} as executable code, the use of an LLM is only required during training, not during inference. Since querying LLMs is both expensive and slow, this is a key advantage of the LangProp approach, which makes integration of LLMs more feasible for real-time applications, such as robotics and autonomous driving.

\section{Experiments}
We demonstrated LangProp's code optimization capability in three domains with increasing complexity. For the LLM, we used GPT 3.5 Turbo 16k model~\citep{chatgpt}.

\subsection{Generalized Sudoku}
A generalized Sudoku puzzle consists of $W \times H$ subblocks, each with $H \times W$ elements, where $H$ and $W$ represent height and width, respectively. A valid solution places numbers from 1 to $WH$ in each cell, such that each row, column and subblock contains no repeated numbers. We trained LangProp on this problem given $100$ samples of unsolved Sudoku puzzles as input and corresponding solutions as output. We define the training objective to be the correctness of the arrived solution, i.e. whether the puzzle completed by a LangProp-learned policy is a valid Sudoku puzzle solution. The setup and update prompts are in \Cref{sec:model_and_prompts}. 
Due to the complexity of the task specification, we found that the LLM queried zero-shot occasionally failed on the first attempt, confusing the task with a standard $3 \times 3$ Sudoku. LangProp filtered out incorrect results during training and identified a fully working solution. Samples of an incorrect zero-shot solution and a correct solution after LangProp training can be found in \Cref{sec:sudoku_solutions}.

\subsection{CartPole}

CartPole~\cite{brockman2016openaigym} is a widely used environment for RL. To make it feasible for LangProp to solve this task, we provided the observation and action specifications, available in the Gymnasium documentation for CartPole-v1. The setup and update prompts are in \Cref{sec:model_and_prompts}. Queried zero-shot, the LLM generated a solution that is simplistic and does not balance the CartPole, achieving a score of 9.9 out of 500. With a simple Monte-Carlo method of optimizing the policy for the total rewards, we obtained improved policies using LangProp, achieving the maximum score of 500.0. Interestingly, LangProp learned a policy that implemented a PID controller to solve the task.

\begin{wrapfigure}{r}{0.5\textwidth}
\includegraphics[width=0.5\textwidth]
{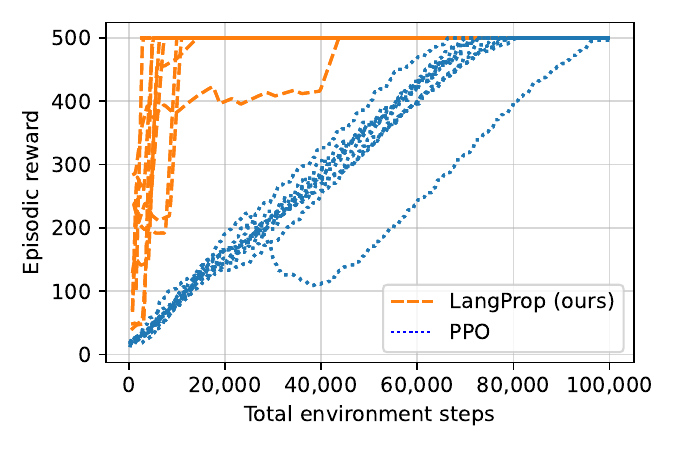}
\caption{The total number of \textit{environment} steps required to learn CartPole-v1 ($10$ seeds per method) in comparison to a RL method (PPO). Most seeds converged to an optimal solution within $10$ LangProp updates.}
\label{fig:langprop_cartpole}
\vspace{-8pt}
\end{wrapfigure}

\Cref{fig:langprop_cartpole} shows learning curves of the LangProp policy for $10$ different seeds. We chose training hyperparameters to be $N^U = N^R = N^K = 3$. Out of $10$ seeds, $9$ converged to an optimal solution within $10$ LangProp updates, and within $10k$ total steps in the CartPole environment. For comparison, we also plotted learning curves of PPO~\cite{schulman2017proximal}, a widely used reinforcement learning algorithm, which converges at around $80k$ environment steps. This shows that certain tasks may be more sample-efficient to solve with LangProp. While it is infeasible to arrive at a correct solution zero-shot, the LangProp optimization loop allows the LLM to discover a correct solution.

Sample results can be found in \cref{sec:cartpole_solutions}. Implementations, prompts, checkpoints, and examples of zero-shot and trained policies are available in the open-sourced repository. 

\subsection{Driving in CARLA}

In this section, we describe how the LangProp framework can be used in the context of autonomous driving in CARLA. 
CARLA~\citep{dosovitskiy17a_carla} is a widely used open-sourced 3D simulator for autonomous driving research, and many prior works on CARLA have open-sourced their expert agents. 
We chose CARLA as a benchmark since (a) autonomous driving requires interpretable driving policies, (b) CARLA has a rich collection of human-implemented expert agents to compare against, and (c) a metric-driven learnable approach would be beneficial since driving decisions are challenging planning problems, and even human-implemented experts have sub-optimal performance. \Cref{sec:related_work_ad} discusses related work on autonomous driving.

\subsubsection{Expert design}
We implemented our expert agent for data collection and to provide action labels to train the LangProp agent with IL. While TransFuser~\citep{chitta2022transfuser} and TF++~\citep{jaeger2023hidden_tfplus} use a computationally expensive 3D bounding box collision detection algorithm, and InterFuser~\citep{shao2023safety_interfuser} uses line collision which is faster but less accurate, we use an efficient polygon collision detection algorithm between ground-projected bounding boxes. 
Safety margins to pedestrians and vehicles are calculated by extrapolating the motion of those actors into the future and checking for any polygon intersections. The target speed is determined to give a $2~s$ margin to any actors in collision course, traffic light and/or stop sign. Steering is evaluated by calculating the angle to a waypoint $4~m$ ahead of the ego vehicle. A PID controller is used for low-level control to convert the target speed and angle to
throttle, brake, and steering. For more details, see \Cref{sec:expert_agent_append}.

\begin{figure}[t]
\includegraphics[width=\textwidth]
{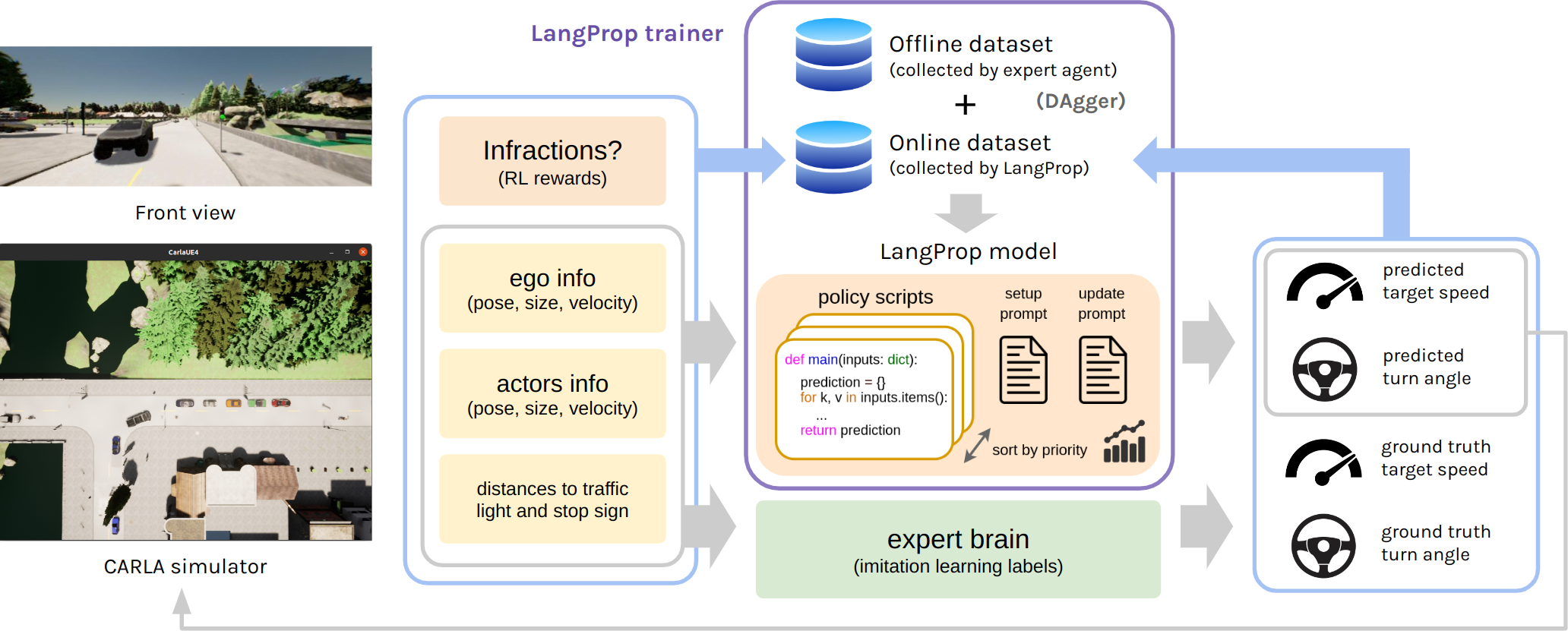}
\caption{An overview of the LangProp agent training pipeline. The LangProp model is updated on a dataset that includes both offline expert data as well as online LangProp data annotated with expert actions, similar to DAgger. The agent is given negative rewards upon infraction.}
\label{fig:langprop_driver}
\vspace{-8pt}
\end{figure}

\subsubsection{LangProp agent}
Similarly to all the baseline experts, we provide privileged information from the CARLA simulator to the agent. Unlike the baseline experts where post-processing is manually implemented, we let LangProp decide how the information should be handled (e.g. converting world coordinates into the ego-centric frame). For the ego vehicle, as well as for all vehicles and pedestrians within a $50~m$ radius, we provide the location (in world coordinates), orientation, speed, length, and width of the actor. Importantly, we do not filter out actors even if they are irrelevant to the driving agent. We also provide the target waypoint ($4~m$ ahead, used by other baseline experts) and the distances to a red traffic light and stop sign along the current lane if they exist. Given this information, the LangProp policy is expected to return a desired speed level ("MOVE": $6~m/s$, "SLOW": $1~m/s$, "STOP": $0~m/s$)\footnote{While it is straightforward for the policy to directly predict the speed or acceleration as numeric values, this makes the task of designing a suitable loss function for IL more challenging and open-ended. Therefore, we opted for a categorical output which simplifies the scoring function.} and a turning angle for the ego vehicle. These are passed to an external PID controller to convert them into throttle, brake, and steering. The function specification in the setup prompt is given in Listing \ref{sec:langprop_drive_setup} as a docstring. Given this function definition, an LLM generates policy script candidates that satisfy the specification and updates them following the procedures in \Cref{sec:langprop}. 

\subsubsection{Imitation Learning, DAgger, and RL}
\label{sec:agent_training}
We explore three major training paradigms often used to train embodied agents - IL, DAgger~\citep{ross2011reduction_dagger}, and RL. In IL, the accuracy of the policy outputs is measured against ground truth expert actions for a pre-collected dataset. IL is known to have issues with out-of-distribution inputs at inference time, since the expert's policy is used to collect the training data, while the learned policy is used for rollouts at inference time. DAgger addresses this issue by labeling newly collected \emph{online} data with expert actions, and adding them to the expert-collected \emph{offline} data to form an aggregate replay buffer. CARLA runs at a frame rate of $20~Hz$. LangProp adds training samples to the replay buffer every $10$ frames, and a batch update is performed after every $100$ new samples.

While DAgger solves the issue of distribution mismatch, the performance of the learned policy is still upper-bounded by the accuracy of the expert. It also does not take into account that certain inaccuracies are more critical than others. In the context of autonomous driving, actions that result in infractions such as collisions should be heavily penalized. Reinforcement learning offers a way of training a policy from reward signals from the environment, which is convenient since we can directly assign penalties upon any infractions according to the CARLA leaderboard~\citep{carla_leaderboard}. 
While RL typically optimizes for maximum returns (discounted sum of future rewards), we simplify the setting by assigning an infraction penalty if there is an infraction in the next $2~s$ window. The agent monitors infractions every 10 frames, and triggers an update upon infractions. 

Since infraction penalties are sparse signals, and will become rarer as the policies improve, we adopt two strategies; (a) we combine RL with IL to provide denser signals, and (b) we sample training data with infractions with $100$ times higher sampling probability.
The expert is only imitated upon no infractions, or if the expert disagrees with the behavior policy which incurred the infraction. An infraction cost is only given when the current policy takes the same action as the behavioral policy that caused the infraction and disagrees with the expert. Details on the objective are in \Cref{sec:driving_objective}.

\subsubsection{Baselines}
We compared the LangProp agent against RL agents with privileged information (Roach~\citep{zhang2021end_carla_roach}, TCP~\citep{wu2022trajectory_tcp}) as well as human-written experts (TransFuser~\citep{chitta2022transfuser}, InterFuser~\citep{shao2023safety_interfuser}, TF++~\citep{jaeger2023hidden_tfplus}, ours). We used the official training and testing routes provided by the CARLA leaderboard~\citep{carla_leaderboard}, as well as the Longest6 benchmark~\citep{chitta2022transfuser} that has longer routes with denser traffic. For the LangProp agent, only the training routes are used for imitation/reinforcement learning at training time, and the saved checkpoints are used for inference during evaluation runs on different routes. 
See \Cref{sec:carla_benchmark} for more details on the benchmark and the routes and towns used. The results are shown in \Cref{table:results}. 

{\renewcommand{\arraystretch}{1.3}
\begin{table*}[t]
\small
\caption{Driving performance of expert drivers in CARLA. The driving score is a product of the route completion percentage $\bar{R}$ and infraction factor $\bar{I}$. DAgger uses both online and offline data.}
\vspace{-8pt}
\begin{center}\small
\begin{tabular}{l ccc ccc ccc}
\toprule
\multirow{2}{1em}{\bf Method} &\multicolumn{3}{c}{\bf Training routes} &\multicolumn{3}{c}{\bf Testing routes} &\multicolumn{3}{c}{\bf Longest6}
\\ 
\cmidrule(rl){2-4} \cmidrule(rl){5-7} \cmidrule(rl){8-10}
&\bf Score $\uparrow$ 
&$\bar{R}\uparrow$
&$\bar{I}\uparrow$
&\bf Score $\uparrow$ 
&$\bar{R}\uparrow$ 
&$\bar{I}\uparrow$
&\bf Score $\uparrow$ 
&$\bar{R}\uparrow$ 
&$\bar{I}\uparrow$
\\ \midrule
Roach expert & 57.8 & 95.9 & 0.61 & 63.4 & 98.8 & 0.64 & 54.9 & 81.7 & 0.67 \\
TCP expert & 64.3 & 92.3 & 0.71 & 72.9 & 93.2 & 0.77 & 46.9 & 63.1 & 0.76 \\
TransFuser expert & 69.8 & 94.5 & 0.74 & 73.1 & 91.3 & 0.80 & 70.8 & 81.2 & 0.88 \\
InterFuser expert & 69.6 & 83.1 & 0.86 & 78.6 & 81.7 & 0.97 & 48.0 & 56.0 & 0.89 \\
TF++ expert & \bf 90.8 & 95.9 & 0.94 & 86.1 & 91.5 & 0.94 & \bf 76.4 & 84.4 & 0.90 \\
\bf Our expert & 88.9 & 92.8 & 0.95 & \bf 95.2 & 98.3 & 0.97 & 72.7 & 78.6 & 0.92 \\
\midrule
LangProp: Offline IL & 0.07 & 0.37 & 0.97 & 0.00 & 0.00 & 1.00 & 0.00 & 0.00 & 1.00 \\
LangProp: DAgger IL & 36.2 & 94.5 & 0.40 & 41.3 & 95.3 & 0.44 & 22.6 & 87.4 & 0.30 \\
LangProp: DAgger IL/RL & 64.2 & 90.0 & 0.72 & 61.2 & 95.2 & 0.64 & 43.7 & 71.1 & 0.65 \\
LangProp: Online IL/RL & \bf 70.3 & 90.5 & 0.78 & \bf 80.9 & 92.0 & 0.89 & \bf 55.0 & 75.7 & 0.73 \\
\bottomrule
\end{tabular}
\end{center}
\label{table:results}
\vspace{-8pt}
\end{table*}
}

\subsubsection{Results}
Our expert and the TF++ expert significantly outperformed all other expert agents in all routes, and our expert outperformed TF++ by a margin on the test routes. The core collision avoidance logic is just 100 lines of code, with additional preprocessing and tooling for data collection. From the breakdown of the scores, our expert seems to prioritize safer driving with fewer infractions (higher infraction factor $\bar{I}$) by trading off route completion compared to TF++ in the Longest6 benchmark.

For the LangProp agent, we observe that training using offline samples, DAgger, and online samples improves performance in this order. Adding the infraction penalties as an additional reinforcement learning objective further improved the performance. The best-performing agent, LangProp trained on online data with IL and RL, achieved better performance than the Roach expert (trained with PPO) as well as the TransFuser and InterFuser experts (both written by researchers) on all benchmarks apart from TransFuser on the Longest6 benchmark. 
Note that TransFuser has an advantage over the Longest6 benchmark since LangProp has never seen this benchmark during training.
The driving policy generated using LangProp is shown in \cref{sec:langprop_driving_policy}.

\begin{figure}[t]
     \centering
     \begin{subfigure}{0.48\textwidth}
         \centering
         \includegraphics[width=\linewidth]{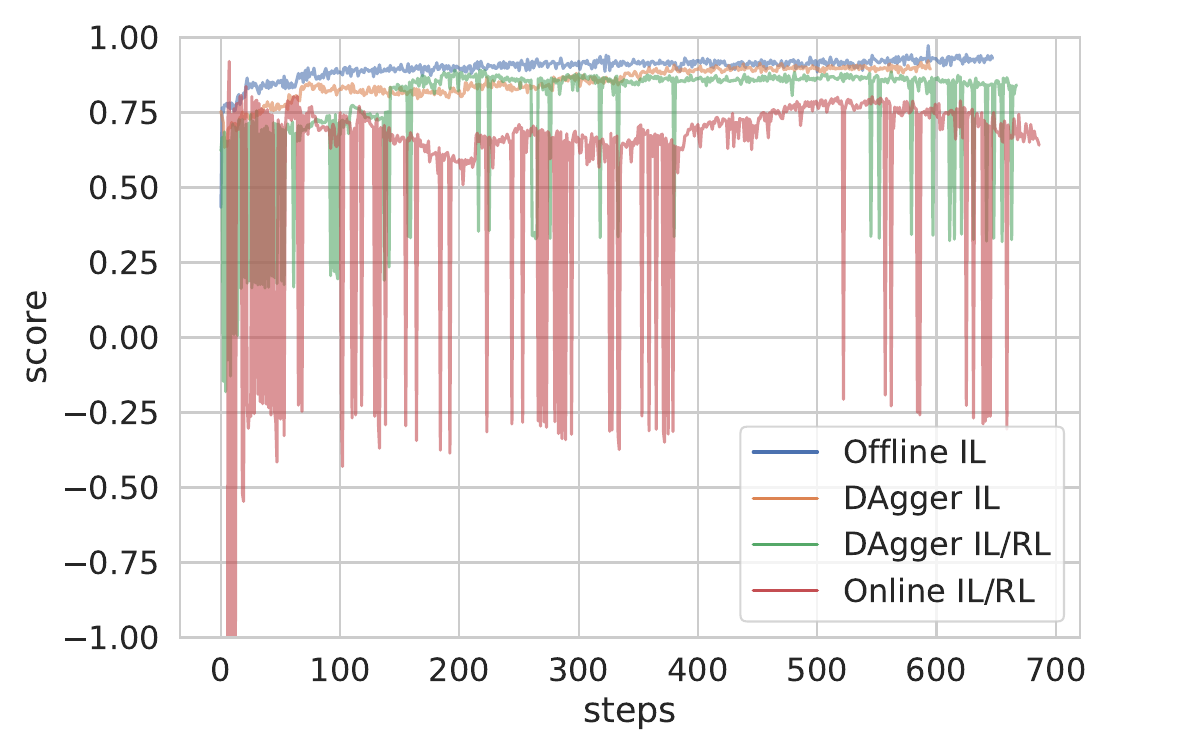}
         \caption{training scores on the replay buffer}
         \label{fig:training_scores}
     \end{subfigure}
     \begin{subfigure}{0.48\textwidth}
         \centering
         \includegraphics[width=\linewidth]{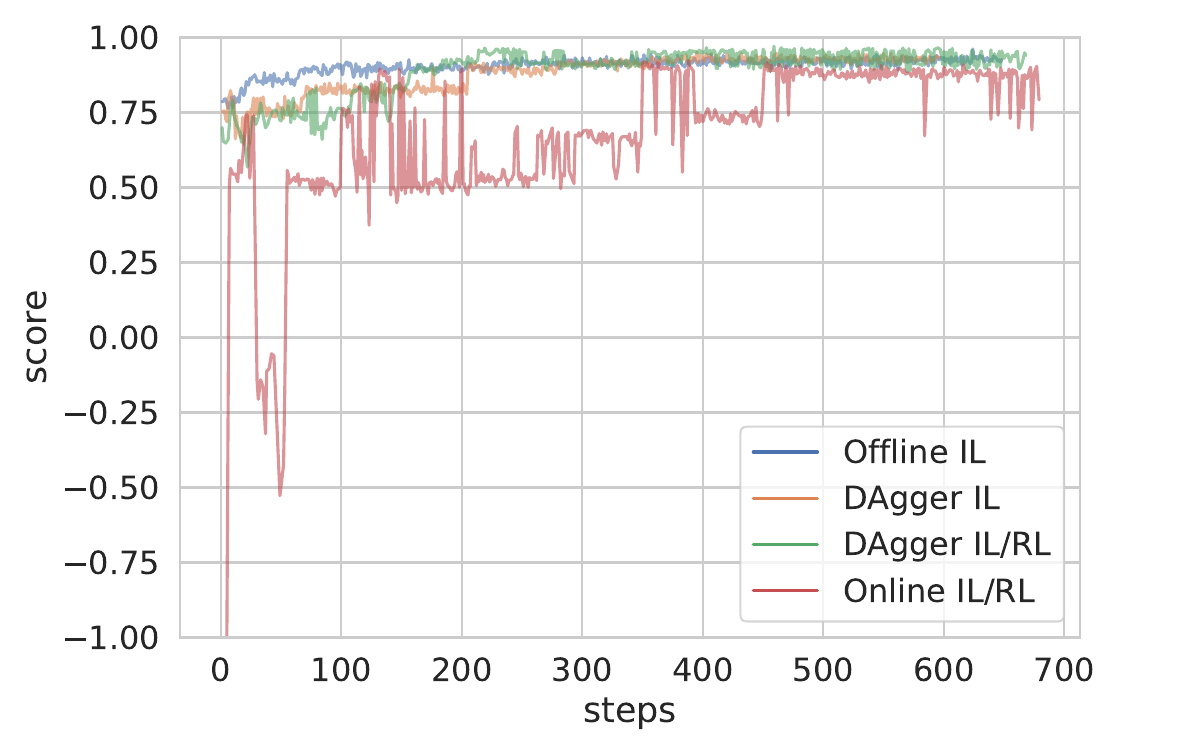}
         \caption{validation scores on the offline dataset}
         \label{fig:val_scores}
     \end{subfigure}
    \caption{Training curves for the different training methods of the LangProp agent. The training scores are evaluated on $1000$ samples from the offline training dataset and/or online replay buffer, and the validation scores are evaluated on $1000$ samples from the offline validation dataset. Updates are performed every $1000$ frames of agent driving, and upon infractions in the RL setting. The score is in the range of $[-10, 1]$ due to exception penalties. We limit the axis to $[-1, 1]$ in the plots.}
    \label{fig:training_curves}
    \vspace{-8pt}
\end{figure}

The result has two important implications. Firstly, the code selection metric (the training objective) plays a large role in the ultimate performance of the code. This is an important finding since prior work on code generation mostly focused on error correction given exceptions. Our results demonstrate that for complex tasks, it is important to treat code generation as an iterative optimization process rather than a zero-shot task. 
Secondly, training using LangProp exhibits similar characteristics as training in deep learning; in deep learning, it is a well-studied problem that policies trained with IL on offline datasets do not generalize to out-of-distribution online data. DAgger and reinforcement learning are two of the common ways of addressing this problem. 
Our results show that these training paradigms can also be effective when used in LangProp. 

\subsubsection{Analysis of training methods}
A common failure mode of offline trained models was that the agent remained stationary indefinitely until the timeout was reached. Upon inspection of the policy code that was generated, we were able to identify the failure to be a phenomenon known as causal confusion in IL~\citep{de2019causal}. A snippet of code responsible for such failure in one of the runs is shown in \Cref{code:causal_confusion}.

This exemplifies the interpretability of LangProp models, allowing us to directly assess the source of failure. The code predicts $0$ speed when the agent's current speed is already close to $0$. Note that this is not a failure of LangProp, but due to such a policy maximizing the IL objective on an offline dataset, bypassing the need to learn a more complex policy. This phenomenon is also common in the context of \textit{deep} IL, and can be avoided by employing training on online data, e.g. DAgger or RL. We believe our work to be the first to report a similar phenomenon using LLMs for policy optimization.

\python{code/causal_confusion.py}{code:causal_confusion}{Causal confusion in offline-trained policy}

The use of online training samples alleviated the issue of causal confusion, leading to selecting policies where the agent has a sensible driving performance. This is because if the agent remains stationary, those samples will accumulate in the replay buffer, resulting in a lower priority for the causally confused policy.
Comparing the results in \Cref{table:results} and the validation scores in \Cref{fig:val_scores}, it seems that the scores on the offline dataset are not indicative of the agent's driving performance. 
From the training scores on the replay buffer and/or offline dataset in \Cref{fig:training_scores}, we see that the agents trained with RL on infractions have spikes corresponding to infractions. This is due to oversampling infractions when they occur, allowing the policy update to immediately address the issue. DAgger has a milder response compared to training just on online data because the offline dataset does not include on-policy infractions. The higher rate of infractions in the training distribution may be why the online trained agent has a lower training score but has a higher driving performance.
 
\section{Conclusion}
\label{sec:conclusion}
We presented LangProp, a framework that uses LLMs for data-driven code optimization, and demonstrated its capability of generating and improving policies in the domains of Sudoku, CartPole and CARLA. In particular, LangProp generated driving policies in CARLA that outperform those that existed when the backbone GPT 3.5 was trained. We showed that classical training paradigms such as IL, DAgger, and RL directly translate to training with LangProp, and the choices of the objective function and the training data distribution can be used to guide which policies are selected. Automatically optimizing the code to maximize a given performance metric has been a key missing feature in few-shot code generation. The LangProp framework provides this feature by reformulating the machine learning training paradigm in the context of using LLMs as code optimizers and treating policy code as parameters of the model. We believe that the LangProp paradigm opens up many possibilities for data-driven machine learning with more interpretability and transparency.

\subsubsection*{Acknowledgments}
The authors would like to thank Oleg Sinavski, Long Chen, Benoît Hanotte, Ana-Maria Marcu, Jan Hünermann and Corina Gurau for insightful research discussions and assistance with compute resources.

\bibliography{iclr2024_conference}

\begin{thebibliography}{63}
\providecommand{\natexlab}[1]{#1}
\providecommand{\url}[1]{\texttt{#1}}
\expandafter\ifx\csname urlstyle\endcsname\relax
  \providecommand{\doi}[1]{doi: #1}\else
  \providecommand{\doi}{doi: \begingroup \urlstyle{rm}\Url}\fi

\bibitem[Abu-Nasser(2017)]{abu2017medical}
Bassem Abu-Nasser.
\newblock Medical expert systems survey.
\newblock \emph{International Journal of Engineering and Information Systems (IJEAIS)}, 1\penalty0 (7):\penalty0 218--224, 2017.

\bibitem[Ahn et~al.(2022)Ahn, Brohan, Brown, Chebotar, Cortes, David, Finn, Fu, Gopalakrishnan, Hausman, et~al.]{ahn2022saycan}
Michael Ahn, Anthony Brohan, Noah Brown, Yevgen Chebotar, Omar Cortes, Byron David, Chelsea Finn, Chuyuan Fu, Keerthana Gopalakrishnan, Karol Hausman, et~al.
\newblock Do as i can, not as i say: Grounding language in robotic affordances.
\newblock \emph{arXiv preprint arXiv:2204.01691}, 2022.

\bibitem[B{\"a}ck \& Schwefel(1993)B{\"a}ck and Schwefel]{back1993overview_evolution}
Thomas B{\"a}ck and Hans-Paul Schwefel.
\newblock An overview of evolutionary algorithms for parameter optimization.
\newblock \emph{Evolutionary computation}, 1\penalty0 (1):\penalty0 1--23, 1993.

\bibitem[Badue et~al.(2021)Badue, Guidolini, Carneiro, Azevedo, Cardoso, Forechi, Jesus, Berriel, Paixao, Mutz, et~al.]{badue2021self}
Claudine Badue, R{\^a}nik Guidolini, Raphael~Vivacqua Carneiro, Pedro Azevedo, Vinicius~B Cardoso, Avelino Forechi, Luan Jesus, Rodrigo Berriel, Thiago~M Paixao, Filipe Mutz, et~al.
\newblock Self-driving cars: A survey.
\newblock \emph{Expert Systems with Applications}, 165:\penalty0 113816, 2021.

\bibitem[Bansal et~al.(2018)Bansal, Krizhevsky, and Ogale]{bansal2018chauffeurnet}
Mayank Bansal, Alex Krizhevsky, and Abhijit Ogale.
\newblock Chauffeurnet: Learning to drive by imitating the best and synthesizing the worst.
\newblock \emph{arXiv preprint arXiv:1812.03079}, 2018.

\bibitem[Bojarski et~al.(2016)Bojarski, Testa, Dworakowski, Firner, Flepp, Goyal, Jackel, Monfort, Muller, Zhang, Zhang, Zhao, and Zieba]{bojarski2016end}
Mariusz Bojarski, Davide~Del Testa, Daniel Dworakowski, Bernhard Firner, Beat Flepp, Prasoon Goyal, Lawrence~D. Jackel, Mathew Monfort, Urs Muller, Jiakai Zhang, Xin Zhang, Jake Zhao, and Karol Zieba.
\newblock End to end learning for self-driving cars, 2016.

\bibitem[Brockman et~al.(2016)Brockman, Cheung, Pettersson, Schneider, Schulman, Tang, and Zaremba]{brockman2016openaigym}
Greg Brockman, Vicki Cheung, Ludwig Pettersson, Jonas Schneider, John Schulman, Jie Tang, and Wojciech Zaremba.
\newblock Openai gym, 2016.

\bibitem[Brown et~al.(2020)Brown, Mann, Ryder, Subbiah, Kaplan, Dhariwal, Neelakantan, Shyam, Sastry, Askell, Agarwal, Herbert-Voss, Krueger, Henighan, Child, Ramesh, Ziegler, Wu, Winter, Hesse, Chen, Sigler, Litwin, Gray, Chess, Clark, Berner, McCandlish, Radford, Sutskever, and Amodei]{NEURIPS2020_gpt3}
Tom Brown, Benjamin Mann, Nick Ryder, Melanie Subbiah, Jared~D Kaplan, Prafulla Dhariwal, Arvind Neelakantan, Pranav Shyam, Girish Sastry, Amanda Askell, Sandhini Agarwal, Ariel Herbert-Voss, Gretchen Krueger, Tom Henighan, Rewon Child, Aditya Ramesh, Daniel Ziegler, Jeffrey Wu, Clemens Winter, Chris Hesse, Mark Chen, Eric Sigler, Mateusz Litwin, Scott Gray, Benjamin Chess, Jack Clark, Christopher Berner, Sam McCandlish, Alec Radford, Ilya Sutskever, and Dario Amodei.
\newblock Language models are few-shot learners.
\newblock In H.~Larochelle, M.~Ranzato, R.~Hadsell, M.F. Balcan, and H.~Lin (eds.), \emph{Advances in Neural Information Processing Systems}, volume~33, pp.\  1877--1901. Curran Associates, Inc., 2020.

\bibitem[CARLA(2020)]{carla_leaderboard}
CARLA.
\newblock Carla autonomous driving leaderboard.
\newblock \url{https://leaderboard.carla.org/}, 2020.

\bibitem[Chen et~al.(2021)Chen, Tworek, Jun, Yuan, de~Oliveira~Pinto, Kaplan, Edwards, Burda, Joseph, Brockman, Ray, Puri, Krueger, Petrov, Khlaaf, Sastry, Mishkin, Chan, Gray, Ryder, Pavlov, Power, Kaiser, Bavarian, Winter, Tillet, Such, Cummings, Plappert, Chantzis, Barnes, Herbert-Voss, Guss, Nichol, Paino, Tezak, Tang, Babuschkin, Balaji, Jain, Saunders, Hesse, Carr, Leike, Achiam, Misra, Morikawa, Radford, Knight, Brundage, Murati, Mayer, Welinder, McGrew, Amodei, McCandlish, Sutskever, and Zaremba]{chen2021evaluating_codex}
Mark Chen, Jerry Tworek, Heewoo Jun, Qiming Yuan, Henrique~Ponde de~Oliveira~Pinto, Jared Kaplan, Harri Edwards, Yuri Burda, Nicholas Joseph, Greg Brockman, Alex Ray, Raul Puri, Gretchen Krueger, Michael Petrov, Heidy Khlaaf, Girish Sastry, Pamela Mishkin, Brooke Chan, Scott Gray, Nick Ryder, Mikhail Pavlov, Alethea Power, Lukasz Kaiser, Mohammad Bavarian, Clemens Winter, Philippe Tillet, Felipe~Petroski Such, Dave Cummings, Matthias Plappert, Fotios Chantzis, Elizabeth Barnes, Ariel Herbert-Voss, William~Hebgen Guss, Alex Nichol, Alex Paino, Nikolas Tezak, Jie Tang, Igor Babuschkin, Suchir Balaji, Shantanu Jain, William Saunders, Christopher Hesse, Andrew~N. Carr, Jan Leike, Josh Achiam, Vedant Misra, Evan Morikawa, Alec Radford, Matthew Knight, Miles Brundage, Mira Murati, Katie Mayer, Peter Welinder, Bob McGrew, Dario Amodei, Sam McCandlish, Ilya Sutskever, and Wojciech Zaremba.
\newblock Evaluating large language models trained on code, 2021.

\bibitem[Chitta et~al.(2022)Chitta, Prakash, Jaeger, Yu, Renz, and Geiger]{chitta2022transfuser}
Kashyap Chitta, Aditya Prakash, Bernhard Jaeger, Zehao Yu, Katrin Renz, and Andreas Geiger.
\newblock Transfuser: Imitation with transformer-based sensor fusion for autonomous driving.
\newblock \emph{IEEE Transactions on Pattern Analysis and Machine Intelligence}, 2022.

\bibitem[Cobbe et~al.(2021)Cobbe, Kosaraju, Bavarian, Chen, Jun, Kaiser, Plappert, Tworek, Hilton, Nakano, et~al.]{cobbe2021training}
Karl Cobbe, Vineet Kosaraju, Mohammad Bavarian, Mark Chen, Heewoo Jun, Lukasz Kaiser, Matthias Plappert, Jerry Tworek, Jacob Hilton, Reiichiro Nakano, et~al.
\newblock Training verifiers to solve math word problems.
\newblock \emph{arXiv preprint arXiv:2110.14168}, 2021.

\bibitem[De~Haan et~al.(2019)De~Haan, Jayaraman, and Levine]{de2019causal}
Pim De~Haan, Dinesh Jayaraman, and Sergey Levine.
\newblock Causal confusion in imitation learning.
\newblock \emph{Advances in Neural Information Processing Systems}, 32, 2019.

\bibitem[D\'idac et~al.(2023)D\'idac, Menon, and Vondrick]{surismenon2023vipergpt}
Sur\'is D\'idac, Sachit Menon, and Carl Vondrick.
\newblock Vipergpt: Visual inference via python execution for reasoning.
\newblock \emph{arXiv preprint arXiv:2303.08128}, 2023.

\bibitem[Dosovitskiy et~al.(2017)Dosovitskiy, Ros, Codevilla, Lopez, and Koltun]{dosovitskiy17a_carla}
Alexey Dosovitskiy, German Ros, Felipe Codevilla, Antonio Lopez, and Vladlen Koltun.
\newblock {CARLA}: {An} open urban driving simulator.
\newblock In Sergey Levine, Vincent Vanhoucke, and Ken Goldberg (eds.), \emph{Proceedings of the 1st Annual Conference on Robot Learning}, volume~78 of \emph{Proceedings of Machine Learning Research}, pp.\  1--16. PMLR, 13--15 Nov 2017.

\bibitem[Driess et~al.(2023)Driess, Xia, Sajjadi, Lynch, Chowdhery, Ichter, Wahid, Tompson, Vuong, Yu, Huang, Chebotar, Sermanet, Duckworth, Levine, Vanhoucke, Hausman, Toussaint, Greff, Zeng, Mordatch, and Florence]{driess2023palme}
Danny Driess, Fei Xia, Mehdi S.~M. Sajjadi, Corey Lynch, Aakanksha Chowdhery, Brian Ichter, Ayzaan Wahid, Jonathan Tompson, Quan Vuong, Tianhe Yu, Wenlong Huang, Yevgen Chebotar, Pierre Sermanet, Daniel Duckworth, Sergey Levine, Vincent Vanhoucke, Karol Hausman, Marc Toussaint, Klaus Greff, Andy Zeng, Igor Mordatch, and Pete Florence.
\newblock Palm-e: An embodied multimodal language model, 2023.

\bibitem[Falcon(2019)]{falcon2019pytorchlightning}
William~A Falcon.
\newblock Pytorch lightning.
\newblock \url{https://github.com/Lightning-AI/lightning}, 2019.

\bibitem[Fried et~al.(2023)Fried, Aghajanyan, Lin, Wang, Wallace, Shi, Zhong, Yih, Zettlemoyer, and Lewis]{fried2022incoder}
Daniel Fried, Armen Aghajanyan, Jessy Lin, Sida Wang, Eric Wallace, Freda Shi, Ruiqi Zhong, Scott Yih, Luke Zettlemoyer, and Mike Lewis.
\newblock Incoder: A generative model for code infilling and synthesis.
\newblock In \emph{The Eleventh International Conference on Learning Representations}, 2023.

\bibitem[Gupta \& Kembhavi(2023)Gupta and Kembhavi]{gupta2023visprog}
Tanmay Gupta and Aniruddha Kembhavi.
\newblock Visual programming: Compositional visual reasoning without training.
\newblock In \emph{Proceedings of the IEEE/CVF Conference on Computer Vision and Pattern Recognition}, pp.\  14953--14962, 2023.

\bibitem[Hayes-Roth(1985)]{hayes1985rule}
Frederick Hayes-Roth.
\newblock Rule-based systems.
\newblock \emph{Communications of the ACM}, 28\penalty0 (9):\penalty0 921--932, 1985.

\bibitem[Hu et~al.(2022)Hu, Corrado, Griffiths, Murez, Gurau, Yeo, Kendall, Cipolla, and Shotton]{hu2022model_mile}
Anthony Hu, Gianluca Corrado, Nicolas Griffiths, Zachary Murez, Corina Gurau, Hudson Yeo, Alex Kendall, Roberto Cipolla, and Jamie Shotton.
\newblock Model-based imitation learning for urban driving.
\newblock \emph{Advances in Neural Information Processing Systems}, 35:\penalty0 20703--20716, 2022.

\bibitem[Huang et~al.(2023)Huang, Xia, Xiao, Chan, Liang, Florence, Zeng, Tompson, Mordatch, Chebotar, et~al.]{huang2023inner_monologue}
Wenlong Huang, Fei Xia, Ted Xiao, Harris Chan, Jacky Liang, Pete Florence, Andy Zeng, Jonathan Tompson, Igor Mordatch, Yevgen Chebotar, et~al.
\newblock Inner monologue: Embodied reasoning through planning with language models.
\newblock In \emph{Conference on Robot Learning}, pp.\  1769--1782. PMLR, 2023.

\bibitem[Jackson(1986)]{jackson1986introduction}
Peter Jackson.
\newblock Introduction to expert systems.
\newblock 1986.

\bibitem[Jaeger et~al.(2023)Jaeger, Chitta, and Geiger]{jaeger2023hidden_tfplus}
Bernhard Jaeger, Kashyap Chitta, and Andreas Geiger.
\newblock Hidden biases of end-to-end driving models.
\newblock \emph{arXiv preprint arXiv:2306.07957}, 2023.

\bibitem[Jiang et~al.(2023)Jiang, Gupta, Zhang, Wang, Dou, Chen, Fei-Fei, Anandkumar, Zhu, and Fan]{jiang2023vima}
Yunfan Jiang, Agrim Gupta, Zichen Zhang, Guanzhi Wang, Yongqiang Dou, Yanjun Chen, Li~Fei-Fei, Anima Anandkumar, Yuke Zhu, and Linxi Fan.
\newblock Vima: General robot manipulation with multimodal prompts.
\newblock In \emph{Fortieth International Conference on Machine Learning}, 2023.

\bibitem[Kendall et~al.(2019)Kendall, Hawke, Janz, Mazur, Reda, Allen, Lam, Bewley, and Shah]{kendall2019learning}
Alex Kendall, Jeffrey Hawke, David Janz, Przemyslaw Mazur, Daniele Reda, John-Mark Allen, Vinh-Dieu Lam, Alex Bewley, and Amar Shah.
\newblock Learning to drive in a day.
\newblock In \emph{2019 International Conference on Robotics and Automation (ICRA)}, pp.\  8248--8254. IEEE, 2019.

\bibitem[Le et~al.(2022)Le, Wang, Gotmare, Savarese, and Hoi]{le2022coderl}
Hung Le, Yue Wang, Akhilesh~Deepak Gotmare, Silvio Savarese, and Steven Chu~Hong Hoi.
\newblock Coderl: Mastering code generation through pretrained models and deep reinforcement learning.
\newblock \emph{Advances in Neural Information Processing Systems}, 35:\penalty0 21314--21328, 2022.

\bibitem[Levinson et~al.(2011)Levinson, Askeland, Becker, Dolson, Held, Kammel, Kolter, Langer, Pink, Pratt, et~al.]{levinson2011towards}
Jesse Levinson, Jake Askeland, Jan Becker, Jennifer Dolson, David Held, Soeren Kammel, J~Zico Kolter, Dirk Langer, Oliver Pink, Vaughan Pratt, et~al.
\newblock Towards fully autonomous driving: Systems and algorithms.
\newblock In \emph{2011 IEEE intelligent vehicles symposium (IV)}, pp.\  163--168. IEEE, 2011.

\bibitem[Li et~al.(2022)Li, Choi, Chung, Kushman, Schrittwieser, Leblond, Eccles, Keeling, Gimeno, Dal~Lago, et~al.]{li2022competition_alphacode}
Yujia Li, David Choi, Junyoung Chung, Nate Kushman, Julian Schrittwieser, R{\'e}mi Leblond, Tom Eccles, James Keeling, Felix Gimeno, Agustin Dal~Lago, et~al.
\newblock Competition-level code generation with alphacode.
\newblock \emph{Science}, 378\penalty0 (6624):\penalty0 1092--1097, 2022.

\bibitem[Liang et~al.(2023)Liang, Huang, Xia, Xu, Hausman, Ichter, Florence, and Zeng]{liang2023codeaspolicies}
Jacky Liang, Wenlong Huang, Fei Xia, Peng Xu, Karol Hausman, Brian Ichter, Pete Florence, and Andy Zeng.
\newblock Code as policies: Language model programs for embodied control.
\newblock In \emph{2023 IEEE International Conference on Robotics and Automation (ICRA)}, pp.\  9493--9500. IEEE, 2023.

\bibitem[Liu et~al.(2023)Liu, Xia, Wang, and Zhang]{liu2023_isyourcodecorrect}
Jiawei Liu, Chunqiu~Steven Xia, Yuyao Wang, and Lingming Zhang.
\newblock Is your code generated by chatgpt really correct? rigorous evaluation of large language models for code generation.
\newblock \emph{arXiv preprint arXiv:2305.01210}, 2023.

\bibitem[Lu et~al.(2022)Lu, Fu, Tucker, Pan, Bronstein, Roelofs, Sapp, White, Faust, Whiteson, et~al.]{lu2022imitation}
Yiren Lu, Justin Fu, George Tucker, Xinlei Pan, Eli Bronstein, Becca Roelofs, Benjamin Sapp, Brandyn White, Aleksandra Faust, Shimon Whiteson, et~al.
\newblock Imitation is not enough: Robustifying imitation with reinforcement learning for challenging driving scenarios.
\newblock \emph{arXiv preprint arXiv:2212.11419}, 2022.

\bibitem[Maddern et~al.(2017)Maddern, Pascoe, Linegar, and Newman]{maddern20171}
Will Maddern, Geoffrey Pascoe, Chris Linegar, and Paul Newman.
\newblock 1 year, 1000 km: The oxford robotcar dataset.
\newblock \emph{The International Journal of Robotics Research}, 36\penalty0 (1):\penalty0 3--15, 2017.

\bibitem[McAllister et~al.(2017)McAllister, Gal, Kendall, Van Der~Wilk, Shah, Cipolla, and Weller]{mcallister2017concrete}
Rowan McAllister, Yarin Gal, Alex Kendall, Mark Van Der~Wilk, Amar Shah, Roberto Cipolla, and Adrian Weller.
\newblock Concrete problems for autonomous vehicle safety: advantages of bayesian deep learning.
\newblock In \emph{Proceedings of the 26th International Joint Conference on Artificial Intelligence}, pp.\  4745--4753, 2017.

\bibitem[Ni et~al.(2023)Ni, Iyer, Radev, Stoyanov, Yih, Wang, and Lin]{pmlr-v202-ni23b-lever}
Ansong Ni, Srini Iyer, Dragomir Radev, Veselin Stoyanov, Wen-Tau Yih, Sida Wang, and Xi~Victoria Lin.
\newblock {LEVER}: Learning to verify language-to-code generation with execution.
\newblock In Andreas Krause, Emma Brunskill, Kyunghyun Cho, Barbara Engelhardt, Sivan Sabato, and Jonathan Scarlett (eds.), \emph{Proceedings of the 40th International Conference on Machine Learning}, volume 202 of \emph{Proceedings of Machine Learning Research}, pp.\  26106--26128. PMLR, 23--29 Jul 2023.

\bibitem[Nijkamp et~al.(2023)Nijkamp, Pang, Hayashi, Tu, Wang, Zhou, Savarese, and Xiong]{nijkamp2022codegen}
Erik Nijkamp, Bo~Pang, Hiroaki Hayashi, Lifu Tu, Huan Wang, Yingbo Zhou, Silvio Savarese, and Caiming Xiong.
\newblock Codegen: An open large language model for code with multi-turn program synthesis.
\newblock In \emph{The Eleventh International Conference on Learning Representations}, 2023.

\bibitem[OpenAI(2022)]{chatgpt}
OpenAI.
\newblock Chatgpt.
\newblock \url{https://openai.com/blog/chatgpt}, 2022.

\bibitem[OpenAI(2023)]{openai2023gpt4}
OpenAI.
\newblock Gpt-4 technical report, 2023.

\bibitem[Ouyang et~al.(2022)Ouyang, Wu, Jiang, Almeida, Wainwright, Mishkin, Zhang, Agarwal, Slama, Ray, Schulman, Hilton, Kelton, Miller, Simens, Askell, Welinder, Christiano, Leike, and Lowe]{NEURIPS2022_b1efde53_instructgpt_human_feedback}
Long Ouyang, Jeffrey Wu, Xu~Jiang, Diogo Almeida, Carroll Wainwright, Pamela Mishkin, Chong Zhang, Sandhini Agarwal, Katarina Slama, Alex Ray, John Schulman, Jacob Hilton, Fraser Kelton, Luke Miller, Maddie Simens, Amanda Askell, Peter Welinder, Paul~F Christiano, Jan Leike, and Ryan Lowe.
\newblock Training language models to follow instructions with human feedback.
\newblock In S.~Koyejo, S.~Mohamed, A.~Agarwal, D.~Belgrave, K.~Cho, and A.~Oh (eds.), \emph{Advances in Neural Information Processing Systems}, volume~35, pp.\  27730--27744. Curran Associates, Inc., 2022.

\bibitem[Paszke et~al.(2019)Paszke, Gross, Massa, Lerer, Bradbury, Chanan, Killeen, Lin, Gimelshein, Antiga, et~al.]{paszke2019pytorch}
Adam Paszke, Sam Gross, Francisco Massa, Adam Lerer, James Bradbury, Gregory Chanan, Trevor Killeen, Zeming Lin, Natalia Gimelshein, Luca Antiga, et~al.
\newblock Pytorch: An imperative style, high-performance deep learning library.
\newblock \emph{Advances in neural information processing systems}, 32, 2019.

\bibitem[Richards(2023)]{autogpt}
Toran~Bruce Richards.
\newblock Auto-gpt.
\newblock \url{https://github.com/Significant-Gravitas/Auto-GPT}, 2023.

\bibitem[Ross et~al.(2011)Ross, Gordon, and Bagnell]{ross2011reduction_dagger}
St{\'e}phane Ross, Geoffrey Gordon, and Drew Bagnell.
\newblock A reduction of imitation learning and structured prediction to no-regret online learning.
\newblock In \emph{Proceedings of the fourteenth international conference on artificial intelligence and statistics}, pp.\  627--635. JMLR Workshop and Conference Proceedings, 2011.

\bibitem[Roziere et~al.(2022)Roziere, Zhang, Charton, Harman, Synnaeve, and Lample]{roziere2021leveraging_unittests}
Baptiste Roziere, Jie Zhang, Francois Charton, Mark Harman, Gabriel Synnaeve, and Guillaume Lample.
\newblock Leveraging automated unit tests for unsupervised code translation.
\newblock In \emph{International Conference on Learning Representations}, 2022.

\bibitem[Sallab et~al.(2017)Sallab, Abdou, Perot, and Yogamani]{sallab2017deep}
Ahmad~EL Sallab, Mohammed Abdou, Etienne Perot, and Senthil Yogamani.
\newblock Deep reinforcement learning framework for autonomous driving.
\newblock \emph{Electronic Imaging}, 29:\penalty0 70--76, 2017.

\bibitem[Schulman et~al.(2017)Schulman, Wolski, Dhariwal, Radford, and Klimov]{schulman2017proximal}
John Schulman, Filip Wolski, Prafulla Dhariwal, Alec Radford, and Oleg Klimov.
\newblock Proximal policy optimization algorithms.
\newblock \emph{arXiv preprint arXiv:1707.06347}, 2017.

\bibitem[Shao et~al.(2023)Shao, Wang, Chen, Li, and Liu]{shao2023safety_interfuser}
Hao Shao, Letian Wang, Ruobing Chen, Hongsheng Li, and Yu~Liu.
\newblock Safety-enhanced autonomous driving using interpretable sensor fusion transformer.
\newblock In \emph{Conference on Robot Learning}, pp.\  726--737. PMLR, 2023.

\bibitem[Singh et~al.(2023)Singh, Blukis, Mousavian, Goyal, Xu, Tremblay, Fox, Thomason, and Garg]{singh2023progprompt}
Ishika Singh, Valts Blukis, Arsalan Mousavian, Ankit Goyal, Danfei Xu, Jonathan Tremblay, Dieter Fox, Jesse Thomason, and Animesh Garg.
\newblock Progprompt: Generating situated robot task plans using large language models.
\newblock In \emph{2023 IEEE International Conference on Robotics and Automation (ICRA)}, pp.\  11523--11530. IEEE, 2023.

\bibitem[Skreta et~al.(2023)Skreta, Yoshikawa, Arellano-Rubach, Ji, Kristensen, Darvish, Aspuru-Guzik, Shkurti, and Garg]{skreta2023errors}
Marta Skreta, Naruki Yoshikawa, Sebastian Arellano-Rubach, Zhi Ji, Lasse~Bj{\o}rn Kristensen, Kourosh Darvish, Al{\'a}n Aspuru-Guzik, Florian Shkurti, and Animesh Garg.
\newblock Errors are useful prompts: Instruction guided task programming with verifier-assisted iterative prompting.
\newblock \emph{arXiv preprint arXiv:2303.14100}, 2023.

\bibitem[Sutton \& Barto(2018)Sutton and Barto]{sutton2018reinforcement}
Richard~S Sutton and Andrew~G Barto.
\newblock \emph{Reinforcement learning: An introduction}.
\newblock 2018.

\bibitem[Touvron et~al.(2023)Touvron, Martin, Stone, Albert, Almahairi, Babaei, Bashlykov, Batra, Bhargava, Bhosale, et~al.]{touvron2023llama}
Hugo Touvron, Louis Martin, Kevin Stone, Peter Albert, Amjad Almahairi, Yasmine Babaei, Nikolay Bashlykov, Soumya Batra, Prajjwal Bhargava, Shruti Bhosale, et~al.
\newblock Llama 2: Open foundation and fine-tuned chat models.
\newblock \emph{arXiv preprint arXiv:2307.09288}, 2023.

\bibitem[Urmson et~al.(2008)Urmson, Anhalt, Bagnell, Baker, Bittner, Clark, Dolan, Duggins, Galatali, Geyer, et~al.]{urmson2008autonomous}
Chris Urmson, Joshua Anhalt, Drew Bagnell, Christopher Baker, Robert Bittner, MN~Clark, John Dolan, Dave Duggins, Tugrul Galatali, Chris Geyer, et~al.
\newblock Autonomous driving in urban environments: Boss and the urban challenge.
\newblock \emph{Journal of field Robotics}, 25\penalty0 (8):\penalty0 425--466, 2008.

\bibitem[Vaswani et~al.(2017)Vaswani, Shazeer, Parmar, Uszkoreit, Jones, Gomez, Kaiser, and Polosukhin]{vaswani2017attention}
Ashish Vaswani, Noam Shazeer, Niki Parmar, Jakob Uszkoreit, Llion Jones, Aidan~N Gomez, {\L}ukasz Kaiser, and Illia Polosukhin.
\newblock Attention is all you need.
\newblock \emph{Advances in neural information processing systems}, 30, 2017.

\bibitem[Wang et~al.(2023)Wang, Xie, Jiang, Mandlekar, Xiao, Zhu, Fan, and Anandkumar]{wang2023voyager}
Guanzhi Wang, Yuqi Xie, Yunfan Jiang, Ajay Mandlekar, Chaowei Xiao, Yuke Zhu, Linxi Fan, and Anima Anandkumar.
\newblock Voyager: An open-ended embodied agent with large language models, 2023.

\bibitem[Wei et~al.(2022)Wei, Wang, Schuurmans, Bosma, Xia, Chi, Le, Zhou, et~al.]{wei2022chainofthought}
Jason Wei, Xuezhi Wang, Dale Schuurmans, Maarten Bosma, Fei Xia, Ed~Chi, Quoc~V Le, Denny Zhou, et~al.
\newblock Chain-of-thought prompting elicits reasoning in large language models.
\newblock \emph{Advances in Neural Information Processing Systems}, 35:\penalty0 24824--24837, 2022.

\bibitem[Wei et~al.(2013)Wei, Snider, Kim, Dolan, Rajkumar, and Litkouhi]{wei2013towards}
Junqing Wei, Jarrod~M Snider, Junsung Kim, John~M Dolan, Raj Rajkumar, and Bakhtiar Litkouhi.
\newblock Towards a viable autonomous driving research platform.
\newblock In \emph{2013 IEEE Intelligent Vehicles Symposium (IV)}, pp.\  763--770. IEEE, 2013.

\bibitem[Wu et~al.(2022)Wu, Jia, Chen, Yan, Li, and Qiao]{wu2022trajectory_tcp}
Penghao Wu, Xiaosong Jia, Li~Chen, Junchi Yan, Hongyang Li, and Yu~Qiao.
\newblock Trajectory-guided control prediction for end-to-end autonomous driving: A simple yet strong baseline.
\newblock \emph{Advances in Neural Information Processing Systems}, 35:\penalty0 6119--6132, 2022.

\bibitem[Xia \& Zhang(2022)Xia and Zhang]{xia2022less_coderepair}
Chunqiu~Steven Xia and Lingming Zhang.
\newblock Less training, more repairing please: revisiting automated program repair via zero-shot learning.
\newblock In \emph{Proceedings of the 30th ACM Joint European Software Engineering Conference and Symposium on the Foundations of Software Engineering}, pp.\  959--971, 2022.

\bibitem[Xia et~al.(2022)Xia, Wei, and Zhang]{xia2022practical_coderepair}
Chunqiu~Steven Xia, Yuxiang Wei, and Lingming Zhang.
\newblock Practical program repair in the era of large pre-trained language models.
\newblock \emph{arXiv preprint arXiv:2210.14179}, 2022.

\bibitem[Xu et~al.(2022)Xu, Alon, Neubig, and Hellendoorn]{xu_polycoder2022}
Frank~F. Xu, Uri Alon, Graham Neubig, and Vincent~Josua Hellendoorn.
\newblock A systematic evaluation of large language models of code.
\newblock In \emph{Proceedings of the 6th ACM SIGPLAN International Symposium on Machine Programming}, MAPS 2022, pp.\  1–10, New York, NY, USA, 2022. Association for Computing Machinery.
\newblock ISBN 9781450392730.
\newblock \doi{10.1145/3520312.3534862}.

\bibitem[Yao et~al.(2023)Yao, Zhao, Yu, Du, Shafran, Narasimhan, and Cao]{yao2022react}
Shunyu Yao, Jeffrey Zhao, Dian Yu, Nan Du, Izhak Shafran, Karthik~R Narasimhan, and Yuan Cao.
\newblock React: Synergizing reasoning and acting in language models.
\newblock In \emph{The Eleventh International Conference on Learning Representations}, 2023.

\bibitem[Yurtsever et~al.(2020)Yurtsever, Lambert, Carballo, and Takeda]{yurtsever2020survey_autonomous}
Ekim Yurtsever, Jacob Lambert, Alexander Carballo, and Kazuya Takeda.
\newblock A survey of autonomous driving: Common practices and emerging technologies.
\newblock \emph{IEEE access}, 8:\penalty0 58443--58469, 2020.

\bibitem[Zhang et~al.(2023)Zhang, Chen, Shen, Ding, Tenenbaum, and Gan]{zhang2022planning}
Shun Zhang, Zhenfang Chen, Yikang Shen, Mingyu Ding, Joshua~B Tenenbaum, and Chuang Gan.
\newblock Planning with large language models for code generation.
\newblock In \emph{The Eleventh International Conference on Learning Representations}, 2023.

\bibitem[Zhang et~al.(2021)Zhang, Liniger, Dai, Yu, and Van~Gool]{zhang2021end_carla_roach}
Zhejun Zhang, Alexander Liniger, Dengxin Dai, Fisher Yu, and Luc Van~Gool.
\newblock End-to-end urban driving by imitating a reinforcement learning coach.
\newblock In \emph{Proceedings of the IEEE/CVF international conference on computer vision}, pp.\  15222--15232, 2021.

\end{thebibliography}
\bibliographystyle{iclr2024_conference}

\appendix
\setcounter{listing}{0}
\renewcommand\thelisting{A.\arabic{listing}}

\appendixpage

\section*{Reproducibility statement}
We open-source the code used in this paper, including code for the general LangProp framework, applying LangProp to tasks such as Sudoku and CartPole, and training and evaluating the LangProp agent in CARLA. More details of the implementation and design decisions can be found in the appendices.

Supplementary materials can be found at \url{https://github.com/shuishida/LangProp/}, which includes the code, pre-trained checkpoints using LangProp, and videos of sample runs by the LangProp agent. 

\section{LangProp model and prompt definitions}
\label{sec:model_and_prompts}
LangProp as a framework can be used to optimize a diverse range of code optimization problems. The functionality of the model is determined by the choices in the setup prompt, the update prompt, and the dataset that the LangProp model is trained on. 

\subsection{Policy setup prompt examples}
\label{sec:setup_prompt_example}
We provide simple examples of learning a Sudoku algorithm, and learning a policy that plays CartPole-v1, a widely used reinforcement learning environment in \cite{brockman2016openaigym}, to show the generality of the framework. The setup prompt should include the specification of the function's inputs and outputs and their types in the form of a docstring.

\prompt{code/sudoku_setup.py}{code:sudoku_setup}{Setup prompt template to learn an algorithm to solve generalized Sudoku}

\prompt{code/cartpole_setup.py}{code:cartpole_setup}{Setup prompt template to learn an agent policy to play CartPole}

\subsection{Policy update prompt example}
\label{sec:update_prompt_example}
The prompt used to update the policy contains the same information as the setup prompt, but in addition, has example inputs and outputs where the code had failed to produce a valid prediction. If there was an exception or printed messages during the execution of the code, this will also be provided as feedback. The LLM is asked to identify the source of the sub-optimal performance and rewrite the code to achieve a higher score.

\prompt{code/sudoku_update.py}{code:sudoku_update}{Update prompt template to learn an algorithm to solve generalized Sudoku}

\prompt{code/cartpole_update.py}{code:cartpole_update}{Update prompt template to learn an agent policy to play CartPole}

\subsection{Model forward pass definition}
The LangProp module captures printed outputs and exceptions and stores them in the policy tracker along with the corresponding inputs during a forward pass. The Python code snippet extracted from the LLM's response and saved as a text string is executed using the \texttt{exec} function in Python. The local scope variables can be accessed via \texttt{locals}.

\python{code/langprop_model_forward.py}{code:model_forward}{Forward passing mechanism of the LangProp module (extract)}

\subsection{Trainer forward-backward definition}
The trainer has a similar abstraction to deep learning training. At every step, it triggers a forward method that calls the policy and stores the inputs, the policy's prediction, and the expected output, and a backward method that updates the policy tracker with the scores, exceptions, or any feedback.

\python{code/langprop_trainer.py}{code:trainer}{Forward-backward pass in the LangProp Trainer (extract)}

\subsection{Policy definition for the LangProp driving agent in CARLA}
\label{sec:langprop_drive_setup}
The driving policy is given the location, orientation, speed, length, and width of the ego vehicle, other vehicles and pedestrians in the scene, the distances to the next red traffic light and stop sign, and the target waypoint ($4~m$ ahead, used by other baseline experts), all in absolute world coordinates.

\python{code/langprop_drive_setup.py}{code:langprop_drive_setup}{Docstring given as part of the setup prompt for the LangProp agent}

\subsection{Notes on specifying the policy}
One of the challenges in the early stages of the project was in specifying the inputs and outputs of the function. Most of the failures in learning a policy were due to misspecification of the inputs, rather than a fundamental problem with the LLM or with LangProp. For instance, we found that it is crucial to specify the units of the input values, e.g. $m/s$, which allowed the LLM to choose sensible values for some internal parameters. It was also important to name input variables explicitly such that it is clear whether the coordinates are given as absolute world coordinates or coordinates relative to the ego vehicle.
A useful property of LangProp is that because the LLM has some understanding of the world from natural language, it can easily incorporate this knowledge when generating the code, constraining the search space of feasible code. We can further guide the LLM to generate policies with certain characteristics, e.g. having a larger safety margin, by expressing our preferences in the prompts. This adds to the benefits of the LangProp approach, where it is easier to encourage policies to exhibit certain behaviors.

\subsection{Details of the prompt template engine}
In the template engine, every line that begins with ``\texttt{\#}" is treated as comments.
Every line that begins with ``\texttt{\$~}" or line blocks in between ``\texttt{\$begin}" and ``\texttt{\$end}" are treated as executable Python code, as well as everything surrounded by \texttt{\{\{ \}\}} in a single line. If a ``\texttt{print}" function is used within the prompt template, it will execute the Python code inside the print function and render the resulting string as a part of the prompt. Variables can be passed to the prompt template engine, and are made accessible in the local scope of the prompt template.

As an example, consider the following prompt template. 

\prompt{code/prompt_template.py}{code:prompt_template}{Example prompt template}

If the prompt template engine is called with the arguments \texttt{\small read\_template("example", people=["Tom", "Jerry"])}, this resolves to: ``Tom and Jerry work here.\symbol{92}nTom is employee No. 1.\symbol{92}nJerry is employee No. 2.".

\subsection{How to choose the priority discount factor}
How the priorities of the policies are calculated has a large effect on the final performance of the trained LangProp model. For a stationary training distribution (e.g. supervised learning on a fixed offline dataset), whether one uses the immediate average score, a running average, or an exponential average does not make a difference except that just using the immediate average score results in a more stochastic result due to fewer numbers of samples. If the computational resources and time are not constrained, one could increase the batch size and just use the immediate average score. If these are constrained, one may adopt a running average with smaller batch sizes. This works when the training distribution is stationary and there are no other changing components other than the policy currently training.

If the training distribution changes or the policy consists of multiple chained modules, each with a learnable sub-policy, we can no longer use a simple running average but have to use either the scores evaluated on a single large batch or the exponential averaging scheme. The current implementation of LangProp does not support multiple chained modules, but is a foreseeable and natural extension to the framework. Changes in the training distribution are expected in DAgger or reinforcement learning. For training our LangProp agent in \Cref{sec:agent_training}, we used a discount factor $\gamma=0$, effectively only using the immediate average scores evaluated on a freshly sampled batch. This is because forward passes through the LangProp driving policies are fast due to not having any complex components so we could afford to have a large batch size. However, in applications where forward passes are expensive and the batch size must be small, using exponential averaging with a non-zero discount factor $\gamma$ is recommended. 

\subsection{Use of the term ``back-propagation"}
\label{sec:backpropagation}
The current LangProp implementation is limited to an update of a single module, i.e. it does not yet accommodate for chaining of modules. 
We have explored this path by making the LLM generate docstrings of helper functions so that submodules can be instantiated, and track priorities also for submodules. However, version tracking of submodules and the mechanism of providing feedback for submodule updates were substantial challenges. 
LangProp v1 does not implement the full back-propagation algorithm, but we refer to a single-layer feedback operation as \emph{back-prop} to highlight the similarities and encourage future research in this area. 

\section{Autonomous Driving and the CARLA benchmark}
\label{sec:related_work_ad}
Approaches to Autonomous Driving can be broadly classified into modular systems and end-to-end systems~\citep{yurtsever2020survey_autonomous}. Most systems take a modular approach~\citep{urmson2008autonomous,levinson2011towards,wei2013towards,maddern20171}, which has human-defined rules that orchestrate separately engineered components for localization and mapping, object detection, tracking, behavior prediction, planning, and vehicle control. Such systems allow compartmentalization and better interpretability, but can be complex and require domain knowledge to maintain and update. Another challenge is error propagation~\citep{mcallister2017concrete}, i.e. the upstream outputs can be erroneous and must be corrected downstream. Recent work has harnessed end-to-end learning to address these issues. Imitation learning (IL)~\citep{bojarski2016end,bansal2018chauffeurnet} optimizes the policy to match actions taken by experts, and is the most widely used approach. However, its performance is upper-bounded by the expert. Deep reinforcement learning has also shown successes in simulation~\citep{sallab2017deep}, on the road~\citep{kendall2019learning}, and in combination with IL~\citep{lu2022imitation}. Our work combines both the benefit of interpretability of expert systems while also taking a data-driven approach, exposing the system to potential failure modes and adverse scenarios during training time and iteratively optimizing the system towards a well-defined driving metric so that the resulting system is robust to adverse events and potential errors in intermediate components.

CARLA~\citep{dosovitskiy17a_carla} is a widely used open-sourced 3D simulator for autonomous driving research, and provides a benchmark~\citep{carla_leaderboard} to evaluate driving performances of both privileged expert agents and sensor-only agents. 
Many prior works on CARLA have open-sourced their expert agents. Roach~\citep{zhang2021end_carla_roach} trained a PPO agent~\citep{schulman2017proximal} on handcrafted reward signals with privileged information. The heavy lifting is done at the reward shaping level, where hazardous agents are identified and the desired speed and pose are computed. Roach expert is also used in MILE~\citep{hu2022model_mile} and TCP~\citep{wu2022trajectory_tcp}, where TCP has an additional emergency braking upon detecting potential collisions. TransFuser~\citep{chitta2022transfuser}, InterFuser~\citep{shao2023safety_interfuser} and TF++~\citep{jaeger2023hidden_tfplus} implement their handcrafted expert systems, either using cuboid intersections or line intersections for hazard detection. TransFuser also introduced the Longest6 benchmark, which consists of longer routes compared to the official CARLA benchmark and is less saturated.

\section{Data collection}
\subsection{Data agent}
To standardize the data collection and evaluation pipeline for both our expert agent and our LangProp agent, we implement a generic \texttt{DataAgent} that collects basic information of the CARLA environment which can be used. These are the 3D bounding box coordinates of the actors in the scene (pedestrians, vehicles, traffic lights, and stop signs), the velocity of the pedestrians and vehicles, distances to the next traffic light and stop sign in the current lane, and the next waypoint to navigate towards. In addition, it also collects the RGB, depth, lidar, segmentation, top-down view, and the expert's control actions which can be used to train image-based driving policies. We created this standardized data collection agent which is decoupled from our expert agent and the LangProp agent, and has the option of turning off sensors that are not used for data collection to save computation time and data storage. 

The data collection agent itself does not have a driving policy. It expects a separate \texttt{AgentBrain} that takes a dictionary of scene information curated by the data agent as input and outputs a vehicle control action (throttle, brake, and steering). All driving agents inherit from the \texttt{DataAgent} class, each with an \texttt{AgentBrain} that implements its driving policy. It is also possible to chain multiple agent brains as an array, where the previous agent brain's control decision is provided as an additional input to the next agent brain. This is useful for our DAgger and online agents, which require expert supervision during online rollouts.

\subsection{Expert agent}
\label{sec:expert_agent_append}
Our expert agent only uses the data collected by the data agent to ensure that the LangProp agent has access to the same privileged information as the expert agent. For every interval of $0.25~s$ up to $2~s$ into the future, we evaluate whether the ego vehicle polygon will intersect any of the actor polygons, assuming that the ego vehicle will maintain velocity, and the other actors will move in the current orientation with a speed less than or equal to the current speed. The ego vehicle polygon is padded forward by $2~m$, and by $2~m$ either left or right upon lane changes. Apart from lane changing, only actors that are ahead of the ego vehicle are considered, i.e. with a field of view of $180\degree$. The traffic light and stop sign that affect the vehicle are identified by querying the associated waypoints in the CARLA simulator. For pedestrians, vehicles, traffic light, and stop sign, the distances to the obstacles are calculated. The normal driving speed is $6~m/s$ ("MOVE"). If any of the distances are reachable within $2~s$ with a $2~m$ margin ("SLOW"), the target speed is set to the speed which allows a $2~s$ margin, and if the distance is below $2~m$ ("STOP"), the target speed is set to $0~m/s$. Steering is evaluated by calculating the angle to the next waypoint, which is $4~m$ ahead of the current position of the ego vehicle. A PID controller is used for low-level control to convert the target speed and angle to throttle, brake, and steering.

\section{Training the LangProp agent}
\subsection{Training strategy}
For all the LangProp agents, the training data is collected only on the training routes in CARLA leaderboard~\citep{carla_leaderboard}, and data collected on the test routes by the expert agent with expert action labels is used as the validation dataset. See \Cref{sec:carla_benchmark} for more details on the routes. For the LangProp agent trained offline, we only use data collected by the expert agent as training data. For the online training, we only use data collected by the current LangProp model's inference policy, i.e. the policy code with the highest priority at the time of rollout. For DAgger training, we have a split of $1000$ training samples collected offline and $1000$ samples collected online in every replay batch to evaluate the objective score. Strictly speaking, DAgger~\citep{ross2011reduction_dagger} should incrementally add new online samples to a buffer initialized with offline samples. However, we found that this prevents the LangProp model from learning from infractions during the early stages of the training, since online samples with infractions are the minority of all the samples. For this reason, we maintained an even split between offline and online samples throughout the training, with a sampling weight of $100$ for samples with infractions. Sampling is without replacements, so that a particular training sample is only sampled once per replay batch.

\subsection{Training objective}
\label{sec:driving_objective}
The training objective for the LangProp driving agent is given as \Cref{eq:drive_objective},

\begin{align}
\begin{split}
\label{eq:drive_objective}
    S(a^\pi, a^{\pi_e}, a^{\pi_b}, I, E) &= \mathbbm{1}\bigl[(a^\pi_{\text{speed}} = a^{\pi_e}_{\text{speed}}) \land [ \lnot I \lor \{ (a^{\pi}_{\text{speed}} \neq a^{\pi_b}_{\text{speed}}) \land (a^{\pi_e}_{\text{speed}} \neq a^{\pi_b}_{\text{speed}}) \} ]\bigr]\\ 
    &+ r_{\text{infrac}} \mathbbm{1}(I \land  (a^{\pi}_{\text{speed}} = a^{\pi_b}_{\text{speed}}) \land (a^{\pi_e}_{\text{speed}} \neq a^{\pi_b}_{\text{speed}}))\\
    &+ r_{\text{angle}} \mathbbm{1}(|a^\pi_{\text{angle}} - a^{\pi_e}_{\text{angle}}| > \theta_{\text{max}}) + r_{\text{error}}\mathbbm{1}(E)
\end{split}
\end{align}
where $a^\pi$, $a^{\pi_e}$ and $a^{\pi_b}$ are actions taken by the current policy, expert policy, and behavior policy used to collect the training sample, respectively, $I$ and $E$ are boolean variables for infraction and exception occurrences, $r_{\text{infrac}} = r_{\text{error}} = r_{\text{angle}} = -10$ are penalties for infraction, exception, and exceeding angle error of $\theta_{\text{max}} = 10\degree$, and $\mathbbm{1}$ equates to 1 if the boolean argument is true, and 0 otherwise. The expert is only imitated when there are no infractions, or if the expert was not the behavior policy that incurred the infraction, and an infraction cost is only given when the current policy takes the same action as the behavioral policy that caused the infraction when the expert chose a different action.

\subsection{Hyperparameters}
Notable training hyperparameters are the number of policies chosen for updates $N^U = 2$, the number of responses per query $N^R = 2$, the number of policies to keep $N^K = 20$, the frequency of batch updates (every $100$ new samples in the replay buffer), batch sizes for online replay data ($1000$) and offline expert data ($1000$), the sampling weight for infractions ($100$), and the infraction, exception, and angle penalties ($r_{\text{infrac}} = r_{\text{error}} = r_{\text{angle}} = -10$). For better performance, it is possible to increase $N^U$, $N^R$, and $N^K$, but with a trade-off of computational time and the cost of using OpenAI API. With our experiment setting, around 700 training steps are taken, 1400 queries are made, and 2800 responses are received from GPT 3.5 per training job, which costs roughly $\$150$.

\section{Evaluation}
\subsection{CARLA benchmark, routes and towns}
\label{sec:carla_benchmark}
The driving scores are computed by the CARLA leaderboard evaluator~\citep{carla_leaderboard}, using the official training and test routes, and the Longest6 benchmark provided by \citet{chitta2022transfuser}. There are towns $1-6$ across the benchmarks. Towns $7-10$ are also used in the official online leaderboard. A breakdown of routes for each benchmark is shown in \Cref{table:carla_towns}. Towns $2$ and $5$ are withheld in the training routes and only appear in the testing routes and the Longest6 benchmark. The Longest6 benchmark has longer routes with denser traffic. 

The main metric of the leaderboard is the driving score, which is evaluated as $\frac{1}{N}\sum_i^N(R_i I_i)$, where $i$ denotes the index of the $N$ routes used for evaluation, $R_i$ is the percentage of route completion of the $i$-th route, and $I_i$ is the infraction factor of the $i$-th route. The infraction factor is a product of infraction coefficients for pedestrian collision ($0.5$), vehicle collision ($0.60$), collision with static objects ($0.65$), running a red light ($0.70$), and running a stop sign ($0.80$). The driving score per route is equal to the route completion $R_i$ when there are no infractions, and is discounted for every infraction by a corresponding infraction factor. Note that in the Longest6 benchmark, the authors decided to remove the stop sign penalty by setting its infraction coefficient to $1.0$, which we adhere to in our experiments.

{
\renewcommand{\arraystretch}{1.2}
\begin{table}[t]
\small
\caption{A breakdown of the number of routes per town, the average length of the routes per town, and traffic density for the training routes, testing routes, and the Longest6 benchmark.}
\begin{center}
\begin{tabular}{l|rrr|rrr|rrr}
\multirow{2}{1em}{\bf Routes} &\multicolumn{3}{c |}{\bf Training routes} &\multicolumn{3}{c |}{\bf Testing routes} &\multicolumn{3}{c}{\bf Longest6}
\\ \cline{2-10} & count & avg. dist. & density & count & avg. dist. & density & count & avg. dist. & density
\\ \hline
Town 1 & 10 & 776.3 & 120   & - & - & 120       & 6 & 898.8 & 500 \\
Town 2 & - & - & 100        & 6 & 911.7 & 100   & 6 & 911.7 & 500 \\
Town 3 & 20 & 1392.5 & 120  & - & - & 120       & 6 & 1797.5 & 500 \\
Town 4 & 10 & 2262.6 & 200  & 10 & 2177.8 & 200 & 6 & 2102.4 & 500 \\
Town 5 & - & - & 120        & 10 & 1230.1 & 120 & 6 & 1444.7 & 500 \\
Town 6 & 10 & 1915.4 & 150  & - & - & 150       & 6 & 2116.7 & 500 \\
\end{tabular}
\end{center}
\label{table:carla_towns}
\vspace{-8pt}
\end{table}
}

\subsection{Software details}
We use CARLA version 0.9.10 for the experiments to maintain consistency with other baseline experts that assume this version. Our expert has been tested both on CARLA version 0.9.10 and version 0.9.11. For LangProp training, we used GPT 3.5 Turbo 16k chat completion API by OpenAI. We used the 16k-token model since the update prompt often exceeds the 8k-token context size of a smaller model. 

\section{Code generation examples}
\label{sec:code_generation_examples}

\subsection{Solutions for Sudoku}
\label{sec:sudoku_solutions}
\subsubsection{Incorrect solution generated zero-shot}

\python{code/sudoku_incorrect.py}{code:sudoku_incorrect}{Example code to solve Sudoku generated zero-shot before LangProp optimization. The code is instructed to solve a general Sudoku with subgrids of size $H \times W$, but confuses it with the standard $3 \times 3$ Sudoku.}

\subsubsection{Correct solution after applying LangProp}

\python{code/sudoku_correct.py}{code:sudoku_correct}{Example code to solve Sudoku after LangProp optimization. The code outputs a correct solution.}

\subsection{Solutions for CartPole}
\label{sec:cartpole_solutions}
\subsubsection{Incorrect solution generated zero-shot}

\python{code/cartpole_zero_shot.py}{code:cartpole_zero_shot}{Example policy to solve CartPole generated zero-shot before LangProp optimization. The overly simplistic policy achieves a mean score of $9.9$ out of $500$.}

\subsubsection{Correct solution after applying LangProp}

\python{code/cartpole_trained.py}{code:cartpole_trained}{Example policy to solve CartPole after LangProp optimization. The policy learns a PID controller and achieves a mean score of $500$ out of $500$.}

\subsection{Driving code generated by LangProp}
\label{sec:langprop_driving_policy}
We show an example driving policy generated using LangProp, trained with both imitation learning and reinforcement learning, as described in \Cref{sec:agent_training}. Please refer to our open-sourced code repository for the full prompts and code used to train the policy, and pre-trained checkpoints for each training setting used in our evaluation.

\python{code/driving_code.py}{code:driving_policy_example}{Example driving policy generated by LangProp, trained with both imitation learning and reinforcement learning.}

\section{Future work}
\label{sec:future_work}
LangProp is a framework that harnesses the capability of LLMs to apply data-driven optimization techniques to code optimization. We do not claim that a solution using LangProp is appropriate for all problems - in fact, neural networks excel in working with continuous state-action spaces and low-level control, whereas LLMs have advantages in handling high-level planning and reasoning tasks, rather than low-level control tasks. Our intention is to propose an alternative learning paradigm that allows LLMs to be used to learn high-level planning which has hitherto been a difficult problem for other machine learning approaches (e.g. neural networks). 

There are numerous future research directions that could improve the capability of LangProp as a training framework, as well as give a better theoretical foundation, such as (a) chaining of modules with a full back-propagation algorithm, (b) improvements to the evolutionary algorithm (e.g. priority mechanism), (c) a robust sampling mechanism for failed examples upon updates, (d) incorporating human feedback in natural language during policy updates, and (e) using LangProp with LLMs fine-tuned for code correction and optimization tasks. In particular, scaling our approach to larger repositories and complex systems would require a multi-modular approach that can propagate useful learning signals to subcomponents if there are multiple failure points in the system. 

Applying LangProp to reinforcement learning tasks has open questions in credit assignment and value estimation. We have demonstrated that reinforcement learning policies written as code can be improved using LangProp if either (a) the policy can be optimized on episodic returns with a Monte-Carlo method (e.g. CartPole), or (b) there is immediate feedback from the environment (e.g. infractions in CARLA). However, for complex tasks that have delayed rewards, it is necessary to have an accurate value/advantage estimator for credit assignment. Since replacing a neural value estimator with a code-based function is not feasible, it is most likely that a hybrid method (having an interpretable code-based actor policy trained with LangProp that uses a value function estimated by a neural network as a critic) would be a way to apply LangProp to complex reinforcement learning scenarios. However, this is also an open-ended question, which calls for further exploration. 

Having an LLM in the RL optimization means that we could potentially harvest more useful signals from the environment, rather than relying just on sparse scalar rewards for updates. For instance, having descriptive feedback from the Gymnasium environment on the failure modes of the agent, given either as a warning or natural language feedback, can significantly accelerate the learning of the RL agent. This also allows a more seamless integration of human-in-the-loop feedback.

Finally, more investigation is required in terms of the robustness and safety of LLM-written applications. This is applicable to all systems that involve code generation. While our framework iteratively improves the quality of the code and filters out potential errors that make the final code policy less likely to contain errors, additional safety mechanisms and firewalls are necessary during the training process, since the code is evaluated based on execution, which could potentially be a source of attacks or risk. We stress the importance of additional safety precautions before deployment.

We believe that LangProp opens up new possibilities for data-driven code development. While zero-shot applications of LLMs have enabled tools such as GitHub Copilot, some suggestions are inaccurate or misaligned with the user’s intentions, whereas if we have data or unit tests that the code needs to satisfy, the code suggestions can be made much more accurate by first running evaluations on these test suites and choosing the best possible suggestion that satisfies the requirements. Planning is one aspect of autonomous driving that has not yet successfully adopted a data-driven approach, for good reasons, since neural networks often struggle to produce generalizable high-level planning rules and are less interpretable. Therefore, most methods currently in deployment have human-engineered planning algorithms. Our LangProp framework is insufficient to replace such systems since it lacks the robustness that human-designed systems have to offer, and more research needs to be done in this direction. We hope that our work will provide inspiration for future research to make the framework more robust and safely deployable in the real world.

\end{document}